\documentclass[]{aastex62}

\usepackage{multirow}
\usepackage{rotating}
\usepackage{graphicx}
\usepackage{epstopdf}
\usepackage{placeins}
\usepackage{color}
\newcommand{\RNum}[1]{\uppercase\expandafter{\romannumeral #1\relax}}

\shorttitle{Superflares of solar-type stars from TESS}
\shortauthors{Tu et al.}

\begin{document}

	\title{Superflares, chromospheric activities and photometric variabilities of solar-type stars from the second-year observation of TESS and spectra of LAMOST}
	
	
	\correspondingauthor{F. Y. Wang}
	\email{fayinwang@nju.edu.cn}
	
	\author[0000-0001-6606-4347]{Zuo-Lin Tu}
	\email{tuzuolin@smail.nju.edu.cn}
	\affil{School of Astronomy and Space Science, Nanjing University, Nanjing 210093, China}
	
	\author[0000-0002-6926-2872]{Ming Yang}
	\affil{School of Astronomy and Space Science, Nanjing University, Nanjing 210093, China}
	\affil{Key Laboratory of Modern Astronomy and Astrophysics (Nanjing University), Ministry of Education, Nanjing 210093, China}
	
	\author[0000-0001-8459-1036]{H.-F. Wang}
	\affil{South$-$Western Institute for Astronomy  Research, Yunnan University, Kunming, 650500, P.\,R.\,China}
	\affil{LAMOST Fellow}
	
	\author[0000-0003-4157-7714]{F. Y. Wang}
	\affil{School of Astronomy and Space Science, Nanjing University, Nanjing 210093, China}
	\affil{Key Laboratory of Modern Astronomy and Astrophysics (Nanjing University), Ministry of Education, Nanjing 210093, China}
	
	
	
	\begin{abstract}
		In this work, 1272 superflares on 311 stars are collected from 22,539 solar-type stars from the second-year observation of Transiting Exoplanet Survey Satellite (TESS), which almost covered the northern hemisphere of the sky. Three superflare stars contain hot Jupiter candidates or ultrashort-period planet candidates. We obtain $\gamma = -1.76\pm 0.11$ of the correlation between flare frequency and flare energy ($dN/dE\propto E^{-\gamma}$) for all superflares and get $\beta=0.42\pm0.01$ of the correlation between superflare duration and energy ($T_{\text {duration }} \propto E^{\beta}$), which supports that a similar mechanism is shared by stellar superflares and solar flares. Stellar photometric variability ($R_{\rm var}$) is estimated for all solar-type stars, and the relation of $E\propto {R_{\rm var}}^{3/2}$ is included. An indicator of chromospheric activity ($S$-index) is obtained by using data from the Large Sky Area Multi-Object Fiber Spectroscopic Telescope (LAMOST) for 7454 solar-type stars. Distributions of these two properties indicate that the Sun is generally less active than superflare stars. We find that saturation-like feature of $R_{\rm var}\sim 0.1$ may be the reason for superflare energy saturating around $10^{36}$ erg. Object TIC 93277807 was captured by the TESS first-year mission and generated the most energetic superflare. This superflare is valuable and unique that can be treated as an extreme event, which may be generated by different mechanisms rather than other superflares.
	\end{abstract}
	
	
	\section{Introduction}
	\label{sec:intro}
	Superflares are energetic phenomena with energies over $10^{33}$ erg, which is much higher than typical solar flares ($\sim 10^{27}$ erg). The study of superflares is important, as they may significantly influence space weather around the star and its hosted planets \citep[e.g.][]{2010AsBio..10..751S, 2016NatGe...9..452A, 2017MNRAS..465L..34A, 2017ApJ...848...41L}. Furthermore, if the Sun can generate superflares, as an amplified version of solar flares, they will definitely put the habitability of the Earth in danger. Before any superflare happens on the Sun, we need to deeply understand this phenomenon at first. Therefore, it seems to be a rational choice to study superflares by using data of other solar-type stars.
	
	From Kepler to the Transiting Exoplanet Survey Satellite
(TESS), we witness the progress of spacetime-domain astronomy. Meanwhile, our understanding about superflares has been changed from debating whether solar-type stars can generate superflares to deeply and statistically understanding superflare uniqueness compared with solar flares. A bunch of works have used Kepler data and found the similarity between solar flares and superflare \citep{2012Natur.485..478M, 2013ApJS..209....5S, 2015EP&S...67...59M}. Until now, it has seemed to be common sense that superflares are not generated through star-planet interactions \citep[e.g.][]{2000ApJ...529.1031R,2004ApJ...602L..53I}, but rather through the magnetic activities \citep[e.g.][]{2013PASJ...65...49S} of the star. Recently, \citet{2021ApJ...906...72O} reused all data from the Kepler main mission and found that the Sun can generate superflares but with a lower rate than those deduced by other literature. In 2018, TESS began to search for exoplanets from the whole sky with 2 yr major missions. Superflares on solar-type stars as by-products of TESS data were collected by \citet{2020ApJ...890...46T} for the first time. Not long after that, \citet{2020MNRAS.494.3596D} collected superflares observed by TESS, and found that there is no correlation between the superflare and rotational phase. However, if we only use photometric data to search for signals of superflares, it will limit our understanding of these superflare stars.
	
	Recently, \citet{2020Sci...368..518R} used photometric variability ($R_{\rm var}$) to find that the Sun is less active than other Sun-like stars. And before that, \citet{2015PASJ...67...33N} and \citet{2016NatCo...711058K} used a similar property called brightness variation amplitude to study superflare stars. Furthermore, chromospheric activity was also estimated in their works using the spectral information. \citet{2015PASJ...67...33N} used H$\alpha$ line and Ca \RNum{2} infrared triplet lines as indicators of stellar chromospheric activity. And \citet{2016NatCo...711058K} used the $S$-index which is an emission measurement of Ca \RNum{2} H and K lines from the Large Sky Area Multi-Object Fiber Spectroscopic Telescope (LAMOST) data. Specifically, they believed that magnetic features (e.g. starspots) on the stellar surface rotate with the star and then cause brightness variation. \citet{2016NatCo...711058K} indirectly proved that superflares were produced from stellar magnetic activities. Recently, \citet{2020ApJ...894L..11Z,2020ApJS..247....9Z} used LAMOST data to statistically analyze magnetic features on G-type stars. They found that Sun-like stars have a relative higher $S$-index than the Sun, which reinforces the conclusion of \citet{2020Sci...368..518R}. 
	
	The above works reveal the capability of using spectral information in the study of stellar activities. \citet{2016NatCo...711058K} used cross-matching results from Kepler and LAMOST, but they have only 11 superflare stars in their database, as Kepler only observed the fixed field of the sky for almost 4 yr.  We expect more superflare stars with LAMOST observations can be found through cross-matching with TESS objects, because TESS changes its observing view every $\sim 27$ days and covers almost the whole sky in 2 yr \citep{2015JATIS...1a4003R}. As the main purpose of our previous work \citep{2020ApJ...890...46T} was to prove and find a usable method for TESS data, we did not include estimations of photospheric variability. Besides, we also did not include evaluation of chromospheric activity before, as we did not found any free resources of spectral survey for the southern hemisphere of the sky. As TESS moved its field of view toward the northern hemisphere of the sky in 2019 July to continue its second-year mission, LAMOST as a ground-based surveying telescope had covered it since 2012 \citep{2012RAA....12.1197C, 2012RAA....12..735D, 2012RAA....12..723Z}. It gives us a wonderful opportunity to statistically analyze superflare stars from both photometric and spectroscopic observations. 
	
	This paper is arranged as follows. In Section \ref{sec:DM}, we just briefly introduce data selection of solar-type stars and superflares. Specifically, we focus on the improvements we made for visual inspection and calculations of $R_{\rm var}$ and $S$-index in this work. In Section \ref{sec:results discuss}, the new findings are all gathered and discussed, especially for the results of the $S$-index and $R_{\rm var}$. Finally, summaries are given in Section \ref{sec:Summary}.

	
\section{Methodology}
\label{sec:DM}
Here we will briefly introduce the method for selecting solar-type stars and superflares. Because some methods are basically the same as those used for TESS first-year observations, one could refer to \citet{2020ApJ...890...46T} for more detailed descriptions. Improvements in this work are specifically introduced in the following.
\subsection{Selection of Solar-type Stars and Superflares}
\label{sec:solar-type data}
In this work, we still use effective temperature ($T_{\mathrm{eff}}$) and surface gravity ($\log g$) as criteria to select solar-type stars. Following the criteria used by other works \citep[e.g.][]{2013ApJS..209....5S,2015EP&S...67...59M,2020ApJ...890...46T}, we also set the effective temperature rang from 5100 to 6000 K and surface gravity $\log g >4.0$. Note that Sun-like stars are also solar-type stars but with stricter criteria as $5600 \mathrm{K} \leqslant
T_{\mathrm{eff}}<6000 \mathrm{K}$ and stellar periods $P>10$ days. Stellar effective temperature and surface gravity are obtained by using the final version of the TESS input catalogue (TIC v8; \citet{2019AJ....158..138S}). Then, 94 stars are excluded from our original data set as these stars are flagged as binary systems, after cross-matching with the Hipparocos-2 catalog \citep{2007A&A...474..653V}. Next, we cross-match the data set with the Gaia Early Data Release 3 (Gaia-EDR3; \citet{2020arXiv201201533G}). Note that stellar effective temperatures of sources in Gaia-EDR3 are all collected from Gaia-DR2 \citep{2018A&A...616A...1G}. Through cross-matching with Gaia-EDR3, we exclude those targets that may be contaminated by other unrelated brighter stars within 21$''$ distance. Those unrelated stars are nontargeting objects in TESS pixel resolution. Because TESS has a pixel resolution of 21$''$, which is much larger than that of Kepler, we should exclude those brighter stars, which may not be distinguished by the TESS cameras and contaminate our targets inside of 1 pixel. For more details, one can refer to  Section 2.1 and Appendix A of  \citet{2020ApJ...890...46T}. We will also briefly discuss the contamination of M dwarfs in section \ref{sec:energy distributions}. After these steps, an additional 209 stars are removed from our database. In total, we select 22,539 solar-type stars for further searching of superflare candidates.

Every single presearch data-conditioned (PDC) light curve of these solar-type stars from TESS has been checked by using the automatic searching algorithm. We have successfully used this method to process TESS first-year observations with 2 minute cadence. This method is obtained by detecting the brightness variations of a star through changes between pairs of consecutive points, which is expressed as
\begin{equation}\label{equ:deltaF}
	\Delta F^{2}\left(t_{i,
		n}\right)=s\left(F_{i}-F_{i-n-1}\right)\left(F_{i+1}-F_{i-n}\right),
\end{equation}
where $i$ stands for the $i$th data point. If $\left(F_{i}-F_{i-n-1}\right) >0$ and $\left(F_{i+1}-F_{i-n}\right)>0$, then $s=1$; otherwise, $s=-1$. Here $n$ describes the number of data points between two pairs of consecutive points. We set $n=2$ and $4$ to search each superflare candidate light curve for two times \citep{2020ApJ...890...46T}. Those points, of which $\Delta F^{2}$ is three times larger than the upper $1\%$ of the $\Delta F^{2}$ distribution, are selected as superflare candidates. Then, we use the quadratic function $F_{q}(t)$ to fit the light curves of these candidates for further calculation of superflare properties. One may refer to Section 2.2 of \citet{2020ApJ...890...46T} for more specific illustrations.

\begin{figure*}[!h]
	\centering
	\includegraphics[width=1\linewidth]{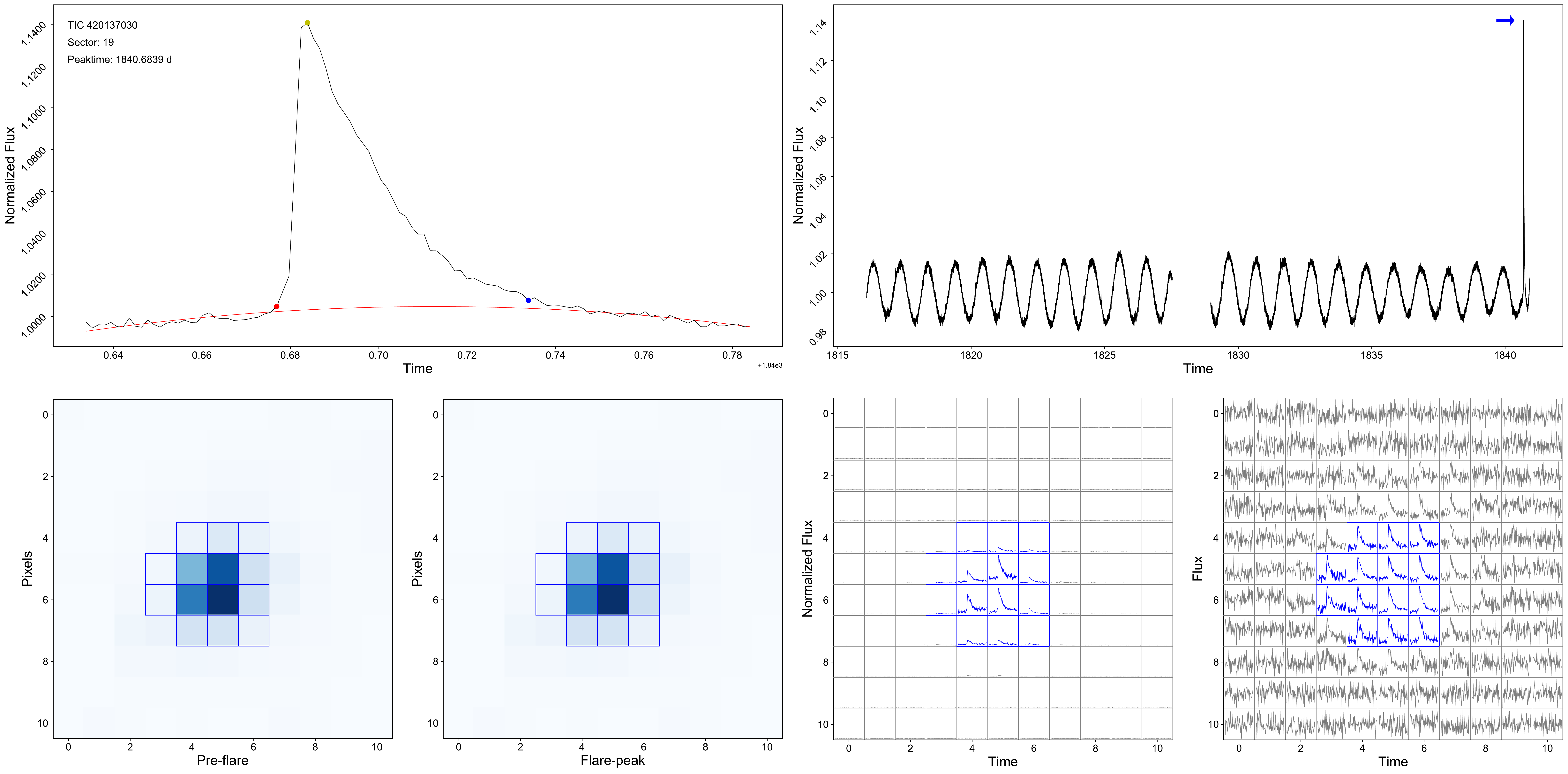}
	\caption{Identification of superflares. The upper left panel shows the superflare light curve in a specific time range. The normalized flux $F(t)$ of this superflare is represented by the black solid line. The red line stands for flux that is fitted by the quadratic function $F_{\rm q}{(t)}$. The start, peak, and end times of the superflare are indicated by red, yellow, and blue dots, respectively. The superflare signal from the whole observing time range of one sector is pointed out in the upper right panel, where the blue arrow marks the peak of this superflare. Pixel-level information is shown in the lower panels, where blue frames in this four panels encircle target aperture masks from the TESS pipeline. The two lower left panels show the pixel images at the preflare and flare peak times, respectively. In these images, the color is positively related to the flux magnitude. The two bottom right panels show the pixel-level light curves of the normalized and unnormalized flux. The light curve in each square frame represents the brightness variation of the corresponding block in the pixel images. The panel containing normalized fluxes shows a standard distribution of light curves described by the PSF. }
	\label{fig:TIC 420137030_19}
\end{figure*}

Comparing with the visual inspection procedure we used for the TESS first-year observation, in this work, we update some details for visually inspecting TESS pixel-level information. Specifically, in Figure \ref{fig:TIC 420137030_19}, compared with our previous work, we add three additional panels to improve the accuracy of the visual inspection. The first one is shown in the upper right panel of Figure \ref{fig:TIC 420137030_19}, which marks the specific moment of the detected superflare candidates in the whole light curve. In the lower panels, we not only show the pixel-level image at the flare peak moment but also attach the image at the preflare time. Therefore, we may easily be sure whether photometric apertures from the TESS pipeline are stable or not. For the pixel-level flux in Figure \ref{fig:TIC 420137030_19}, we also attach the normalized flux for each corresponding pixel. This improvement reduces noises from other unrelated pixels and bolsters superflare signals. From this panel, it is obvious that the results can be described by the point-spread function (PSF). And it is more efficient to confirm those superflare candidates, which show the standard shape of rapidly rising and slowly decaying, as shown in the the upper left panel of Figure \ref{fig:TIC 420137030_19}. Through visual inspections, we exclude many false-positive candidates 
and retain 1272 superflares for further research.

\subsection{Estimation of Superflare Properties and Stellar Periodicity}
\label{sec:properties of superflare}
We estimate stellar luminosity through the Stefan–Boltzmann law 
\begin{equation}\label{equ:lumi}
	L_{*}=4 \pi R_{*}^{2} \sigma_{\mathrm{sb}} T_{*}^{4},
\end{equation}
where the surface effective temperature and stellar radius are denoted as $T_{*}$ and $R_{*}$, and $\sigma_{\mathrm{sb}} $ is the Stefan–Boltzmann constant. The superflare energy is calculated by 
\begin{equation}\label{equ:flare_energy}
	E_{\text {flare }}=\int L_{*} F_{\text {flare }}(t) d t,
\end{equation}
where $ F_{\text {flare }}$ is calculated by
\begin{equation}\label{equ:flare_flux}
	F_{\text {flare }}(t) = F(t)-F_{\rm q}{(t)}.
\end{equation}
Here $F_{\rm q}{(t)}$ is the flux fitted by quadratic function. The start and end times of each superflare are the first and last points, of which three times the photometric errors is less than $F_{\rm flare}(t)$,
\begin{equation}\label{equ:errflux}
	3\times F_{\rm error}(t) < F_{\rm flare}(t),
\end{equation}
where $F_{\rm error}(t)$ is the photometric error, which is provided by the TESS pipeline. 
Superflare duration is defined as the time interval between the start time and end time, which is also the integrating range of Equation (\ref{equ:flare_energy}).
As in our previous work, we also use the Lomb-Scargle method to estimate stellar periods \citep{1976Ap&SS..39..447L,1982ApJ...263..835S}. The false-alarm probability is set to be $10^{-4}$ to search the most probable period for a star \citep[e.g.][]{2019MNRAS.489.5513C}. 

\subsection{Photometric Variability}\label{sec:rvar}
Many works have also used different methods to measure the amplitude of the up-down oscillation in the light curve. For example, \citet{2015ApJS..221...18H,2018ApJS..236....7H} used the rms of the stellar fluxes to estimate the effective fluctuation range of a light curve. The amplitude range of light curve can be understood as surface magnetic features rotating with the star \citep[e.g.][]{2020ApJ...901L..12I, 2020A&A...633A..32S}. In this work, we denote the photometric variability range as $R_{\rm var}$. We use the 95\% and 5\% ranked normalized flux as up and down limits at first \citep{2010ApJ...713L.155B}. Then, the $R_{\rm var}$ is deduced by calculating the difference of the two limits as

\begin{equation}
	\label{equ:rvar}
	R_{\rm var} = F_{95\%} - F_{5\%},
\end{equation}
	
where $F$ stands for the ranked normalized flux of the whole light curves of the star. Note that, in some literature, $R_{\rm var}$ values are multiplied by 100 and in units of percentile, which we do not practice in this work. For all 22,539 stars, $R_{\rm var}$ is calculated through the corresponding light curves.

\subsection{LAMOST Data and Stellar Chromospheric Activity}\label{sec:sindex}
LAMOST has been surveying the sky since 2012, mainly in the northern hemisphere of the sky \citep{2012RAA....12.1197C,2012RAA....12..735D,2012RAA....12..723Z}. It released over 10 millions low-resolution spectra (LRS) with wavelengths from 3700 to 9000 Å. In 2018, LAMOST started a new 5 yr medium-resolution spectroscopic  (MRS survey. And in 2019, the LAMOST DR6 was released and provided MRS spectra that cover the wavelength range from 4950 to 5350 Å in the blue arm and 6300 to 6800 Å in the red arm \citep{2020arXiv200507210L}. We use LRS data from 3890-4020 Å of LAMOST DR7 v1.1 and DR8 v0 in this work. The DR7 v1.1 covers all LRS data from 2012 to 2018, and DR8 v0 only covers data observed since 2019.\footnote{http://dr.lamost.org/data-release.html}

We cross-match all 22,539 solar-type stars in our database with these two LAMOST data releases. We only keep those spectra according to three criteria. (1) The signal-to-noise ratio (S/N) of the spectral blue part should be larger than 10 to ensure accuracy of further analysis. (2) The distance between the targets of TESS and corresponding spectrum of LAMOST should be less than 1$''$. (3) As there might be more than one spectrum matching one TESS target, we only adopt the spectrum that is the nearest one around the target and the latest one observed. In total, 7454 solar-type stars have matched LAMOST spectra, of which 79 stars have superflares. Spectral data are now  available for further estimation of chromospheric activity.

\citet{1913ApJ....38..292E} found a sharp reversal at the K line of Ca from those stellar spectra, which were similar to the Sun. Then, \citet{1968ApJ...153..221W} used photometric results, which showed reliability for using Ca \RNum{2} H and K lines for indicating stellar chromospheric activity. To measure these two lines, \citet{1978PASP...90..267V} defined a method to calculate the mean flux of the H and K lines. This method laid foundation for the best-known $S$-index, a reliable indicator of stellar chromospheric activity. In this work, we also use the LRS from LAMOST observations. Following \citet{2016NatCo...711058K}, we use the reformed version of the $S$-index, which is expressed as 
\begin{equation}\label{equ:sindex}
	S=\alpha \cdot 8 \cdot \frac{\Delta \lambda_{\mathrm{HK}}}{\Delta \lambda_{\mathrm{RV}}} \cdot \frac{\tilde{H}+\tilde{K}}{\tilde{R}+\tilde{V}}=\alpha \cdot 8 \cdot \frac{1.09 {\rm Å}}{20 {\rm Å}} \cdot \frac{\tilde{H}+\tilde{K}}{\tilde{R}+\tilde{V}},
\end{equation}	
where 8 is the ratio of exposure time between channels of H and K lines and other referenced channels of the Mount Wilson spectral photometer \citep{2016NatCo...711058K}. Specifically, $\tilde{H}$ and $\tilde{K}$ are the mean fluxes of the triangle bandpass with $1.09\dot{\mathrm{A}}$ FWHM, so $\Delta \lambda_{\mathrm{HK}} =1.09 {\rm Å}$. These band-passes center at 3968 and 3934 Å. $\tilde{R}$ and $\tilde{V}$ are mean fluxes of bandpasses centered at 3901 and 4001 Å, respectively. These two bandpasses are with 20 Å width, so $\Delta \lambda_{{RV}} =20 {\rm Å}$. \citet{2016NatCo...711058K} proved that by taking $\alpha = 1.8$ for the spectrum from LAMOST, their $S$-indexes have an identical distribution to that of the Isaacson and Fischer sample \citep{2010ApJ...725..875I}. So, we also take the normalization constant $\alpha = 1.8$ from \citet{2007AJ....133..862H}. We use  $\log \sigma(S)=-\log {\rm S / N}-0.5$ to roughly estimate the uncertainty of the $S$-index, where ${\rm S / N}$ is the signal-to-noise ratio of the spectral blue part. This relation is obtained through $\chi^{2}$ minimization in the fitting relation between standard deviation of the $S$-index and ${\rm S / N}$ \citep{2016NatCo...711058K}.

\section{Results and discussions}
\label{sec:results discuss}
After practicing all of above procedures, stellar properties are presented in Tables \ref{tab:non-flare stars} and Table \ref{tab:flarestars}, for nonflare stars and flare stars, respectively. We have searched 1272 superflares through visually checking, as shown in Figure \ref{fig:TIC 420137030_19}. The properties of these superflares are listed in Table \ref{tab:Flares}.
\subsection{Energy Distributions}
\label{sec:energy distributions}
\begin{figure*}[!h]
	\centering
	\includegraphics[width=1\linewidth]{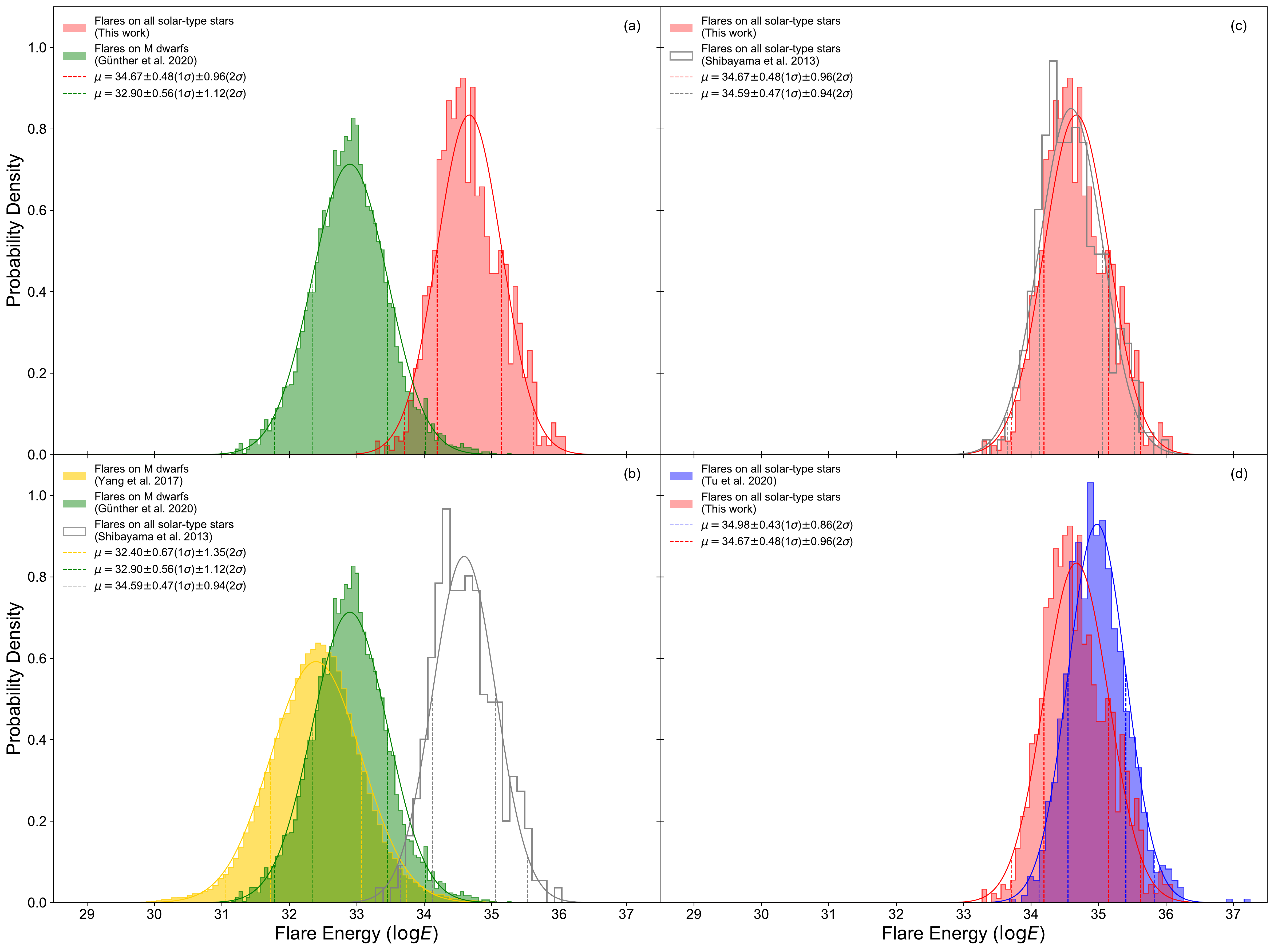}
	\caption{Energy distributions of stellar flares on M dwarfs and superflares on solar-type stars. Each kind of histogram is fitted by the normal distribution (Equation (\ref{equ:normal distribution})). In panels (a) and (b), stellar flares on M dwarfs are represented by green for TESS observations \citep{2020AJ....159...60G} and yellow for Kepler observations \citep{2017ApJ...849...36Y}. In panels (c) and (d), the red and blue histograms represent distributions from superflares on solar-type stars observed by TESS from this work and \citet{2020ApJ...890...46T}, respectively. For comparison, the superflares observed by Kepler \citep{2013ApJS..209....5S} are represented by gray histograms in panels (b) and (c). For each distribution, solid colored curves represent normal distribution fitting of $\log E_{\rm flare}$. The $1\sigma$ and $2\sigma$ standard deviations are shown by dashed lines, and their values are listed in the legend of each panel. 
	}\label{fig:Mflare}
\end{figure*}

We gather all superflare energy distributions in Figure \ref{fig:Mflare}, including M dwarf flares observed by TESS \citep{2020AJ....159...60G} and Kepler \citep{2017ApJ...849...36Y}. For superflares on solar-type stars, we import superflares from Kepler \citep{2013ApJS..209....5S} and TESS first-year observations \citep{2020ApJ...890...46T}. The normal distribution has been fitted for $E_{\rm flare}$ in log scale. The probability density function of the normal distribution is expressed as
\begin{equation}\label{equ:normal distribution}
	f(x)=\frac{1}{\sigma \sqrt{2 \pi}} e^{-\frac{1}{2}\left(\frac{x-\mu}{\sigma}\right)^{2}},
\end{equation}
where $\mu$ is the expectation of this distribution, and $\sigma$ is the standard deviation. 

We have all solar-type stars in this work cross-matched with Gaia-EDR3 to select those TESS targets that may contain M dwarf candidates. However, they are not excluded from our database for the following reasons. (1) The M dwarfs are stars with 3000 K $<T_{eff}<$ 4000 K, and $\log g>$4.0 \citep[e.g.][]{2017ApJ...849...36Y}. However, the stellar parameter of the surface gravity ($\log g$) is not included in the Gaia-EDR3 catalog. So, those selected TESS targets only contain M dwarf candidates. We keep them in our database and mark them with flags, as shown in Tables \ref{tab:non-flare stars} and \ref{tab:flarestars}. (2) As Kepler has better pixel resolution than TESS, in Figure \ref{fig:Mflare}(b), we compare the energy distributions of superflares on solar-type stars \citep{2013ApJS..209....5S} and M dwarfs \citep{2017ApJ...849...36Y} from Kepler observations. Apparently, superflares on solar-type stars (gray histograms) basically have energies over 2 orders of magnitude larger than those of M dwarfs (yellow histograms). As shown in panel (a) of Figure \ref{fig:Mflare}, the superflares of this work still apparently show higher energies than the superflares on M dwarfs observed by TESS \citep{2020AJ....159...60G}. This result indicates the low possibility for flares on M dwarfs contaminating our superflare database, even if they generate flares more actively than solar-type stars. (3) Here is the most important reason. We use a database of superflares on M dwarfs from the first 2 months of the TESS mission \citep{2020AJ....159...60G}. The flare peak luminosity of 5398 superflares on M dwarfs is calculated using 
\begin{equation}\label{equ:normal distribution}
L_{\rm {peak}}=L_{*}\cdot F_{\rm {flare }}(t_{\rm peak}),
\end{equation}
where $F_{\rm {flare }}(t_{\rm peak})$ is the amplitude at the peak moment of the superflare on M dwarfs. The largest $L_{\rm {peak}}$ is equal to $7.22 \times 10^{32} \, {\rm erg\,s^{-1}}$  from TIC 260506296, which was also specifically shown in the Figure 10 of \citet{2020AJ....159...60G}. In our database, the lowest quiescent stellar luminosity ($L_{*}$) of flare stars is equals to $9.67 \times 10^{32} \,{\rm erg\,s^{-1}}$, which is higher than the largest flare peak luminosity ($L_{\rm {peak}}$) of M dwarfs. And only four nonflaring stars in our database have smaller stellar luminosities than the largest $L_{\rm {peak}}$ of the M dwarfs. In short, the superflares on M dwarfs may not affect the observation of solar-type stars, even if they cannot be distinguished in just 1 pixel. Therefore, the possibility that our database is contaminated by superflares on M dwarfs is very low. (4) As TESS has 21$''$ pixel resolution, we may not be able to distinguish M dwarf candidates apart from the main target. Statistically, only 7.23\% (92 of 1272) superflares are from 9.97\% (31 of 311) flare stars the may contain M dwarfs within 21$''$ radius. Specifically, all M dwarf candidates of these 31 flare stars are 1.5 mag fainter (0.25 flux of the Gaia band) than the corresponding flare star. Besides, through cross-matching with LAMOST data, none of these 31 flare stars have M dwarf candidates within 21$''$ radius. As these superflares do not occupy the main part of the whole database, they could not significantly influence the statistical results in the following sections.

As the bandpass filter of TESS (600-1000 nm) is less sensitive to shorter wavelengths than Kepler filter (420-900 nm), TESS is less capable of detecting relatively small flares than Kepler do \citep{2019MNRAS.489..437D,2020ApJ...890...46T}. We also get a similar conclusion from Figure \ref{fig:Mflare}(b). 
 In this work, as shown in Figures \ref{fig:Mflare}(c) and (d), the data of the TESS second-year observation (red histograms) get $\mu \sim 34.67$ which is 0.08 higher than the $\mu \sim 34.59$ of the Kepler data (gray histograms), but 0.31 less than the $\mu \sim 34.98$ of the TESS first-year observation (blue histograms). However, TESS observes superflares with slightly higher energies than those of Kepler data. The difference of $\mu$ between superflares from TESS and Kepler is reduced from 0.39 (TESS first-year observation) to 0.08 (TESS second-year observation), which may be a result of the improved accuracy of visual inspection in this work, as described in Section \ref{sec:solar-type data} and Figure \ref{fig:TIC 420137030_19}. The improvement bolsters flare signals by reducing noises from other unrelated pixels and makes those relatively less energetic superflares more easily to be seen. Also, TESS has upgraded their data processing pipeline.\footnote{https://heasarc.gsfc.nasa.gov/docs/tess/reprocessed-tess-data-coming-soon.html} In this work, we use the reprocessed data for sectors 14-19 of TESS. Besides, in Section \ref{sec:frequencies}, the distribution of flare peak amplitudes shown in Figure \ref{fig:oc_fre}(a) also indicates that the superflares in our database have slightly lower peak amplitudes than those of previous works, which pulls down the energy distribution of superflares in this work. Furthermore, in Section \ref{sec:sindex and rvar}, the distribution of $R_{\rm var}$ of this work (red histograms) shown in Figure \ref{fig:allRvar}(c) has an overall lower $R_{\rm var}$ than that of our previous works (blue histograms). From the scatter plot of Figure \ref{fig:allRvar}(d), we may find a positive relation between maximum flare energy and $R_{\rm var}$. The overall lower $R_{\rm var}$ may also be the reason why the superflare energy in this work is slightly lower than that of \citet{2020ApJ...890...46T}. 

\subsection{Stellar Periodicity Distributions}
\label{sec:period distributions}
\begin{figure*}[!h]
	\centering
	\includegraphics[width=1\linewidth]{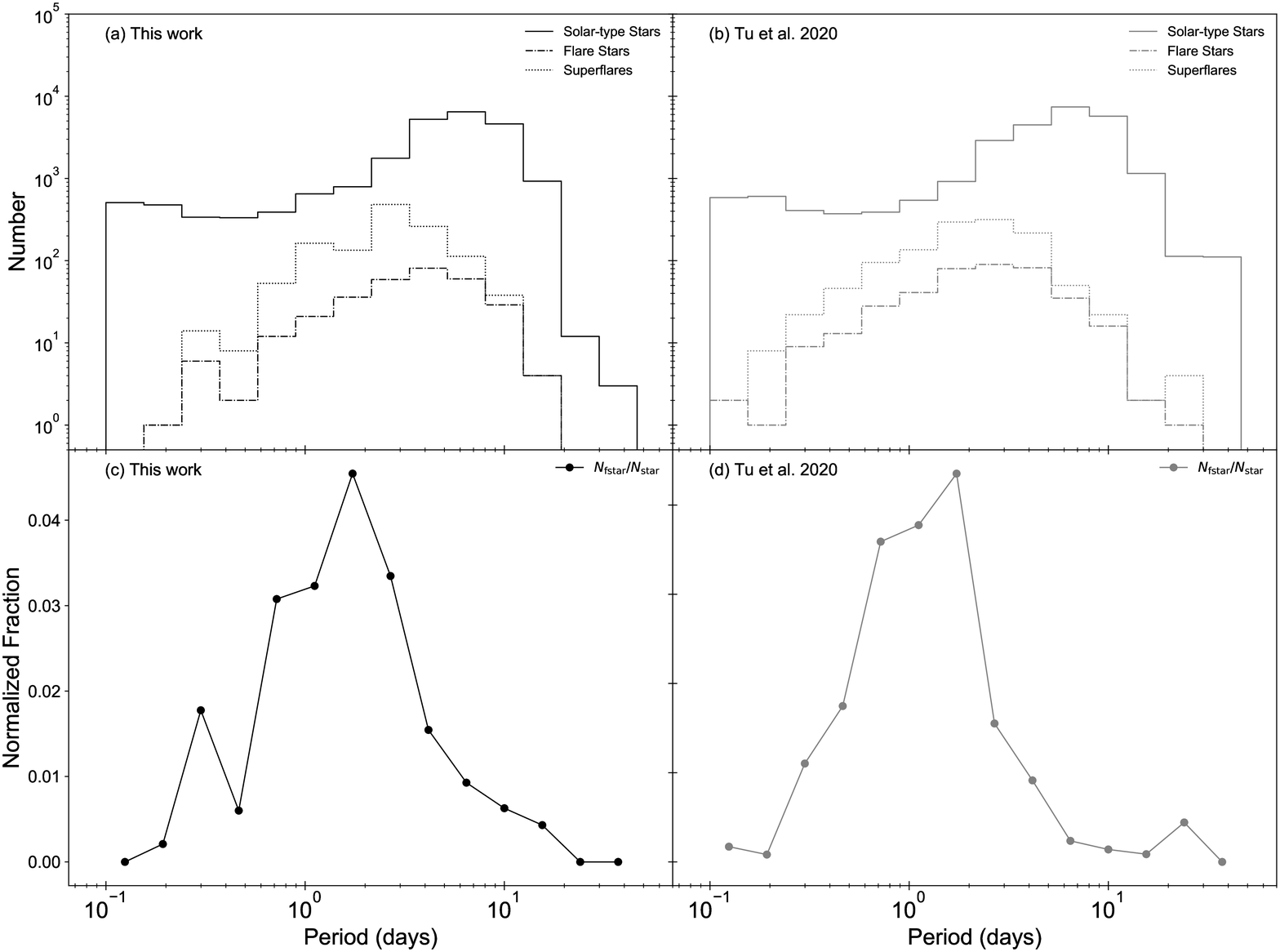}
	\caption{Distribution of stellar periods. The solid, dashed-dotted and dotted lines represent counts of solar-type stars, flare stars, and superflares, respectively. Panels (c) and (d) show the fraction of flare stars over solar-type stars as a function of period. The results of this work are shown in panels (a) and (c), and specific numbers are listed in Table \ref{tab:num_period}. The results of \citet{2020ApJ...890...46T} are shown in panels (b) and (d) for comparison. 
	}
	\label{fig:num_period}
\end{figure*}

Table \ref{tab:num_period} lists counts of solar-type stars, superflares, and flare stars in each period bin. More intuitively, period distributions are shown in Figure \ref{fig:num_period}(a). It shows that the period distributions are not changed significantly from the TESS first-year observation, as shown in Figure \ref{fig:num_period}(b). For each period bin of Figures \ref{fig:num_period}(a) and (b), the corresponding normalized flare star fraction (${N_{\rm fstar}}/{N_{\rm star}}$) is shown in panels (c) and (d). It is evident that the coverage of star spots is negatively related with stellar periodicity (e.g. Figure 15 of \citet{2019ApJ...876...58N}, and Figure 6 of \citet{2021ApJ...906...72O}). In other words, stellar magnetic activity is negatively related to stellar period. From panels (c) and (d) of Figure \ref{fig:num_period}, we also see the decreasing trend of the fraction of flare stars with period $P > 10\; { \rm days}$. So it might be the evidence that magnetic features like star spots are key to generate superflares \citep{2013PASJ...65...49S}. For those stars with period $P > 10\; { \rm days}$, the possibility of producing large magnetic features (e.g. star spots) is low, which may cause lower detectability of their stellar periods. So, the number of stars with period $P > 10\; { \rm days}$ decreases dramatically, as shown in panels (a) and (b) of Figure \ref{fig:num_period}.

As introduced in Section \ref{sec:properties of superflare}, we use the Lomb-Scargle method to estimate stellar periodicity, which is mainly dependent on the brightness variation of stellar light curves. Note that this method may be biased toward selecting stars with large star spots that cause a brightness variation of the stellar light curves. For stars without detectable variability, the periodicity may not be accurately estimated compared with the spectral method. So, it is possible that the real fraction of rapidly and slowly rotating stars in the observed fields may be different from that deduced from the Lomb-Scargle period estimation \citep{2019ApJ...876...58N}. We refer to the result from Table 8 of \citet{2020ApJ...890...46T}, which estimated the number of fractions through the empirical gyrochronology relation. Honestly, this relation is empirical and not totally reliable \citep[e.g.][]{2015A&A...577L...3T,2016Natur.529..181V}. Appendix B of \citet{2020ApJ...890...46T} discussed this relation in detail; one could refer to it for more information. In order to compare, we set four data sets according to stellar effective temperatures and periods. The related results from \citet{2019ApJ...876...58N}, \citet{2020ApJ...890...46T}, and this work are all collected in Table \ref{tab:N10result}. From this table, we find that the fraction of $N_{\rm star}(P > 10\; { \rm days})/N_{\rm all}$ in this work is basically the same as that from the TESS first-year observation \citep{2020ApJ...890...46T}. After comparing our results with those deduced by the gyrochronology relation and \citet{2019ApJ...876...58N}, we still have a lower fraction of $N_{\rm star}(P > 10\; { \rm days})/N_{\rm all}$ than theirs. However, this will not significantly affect the following results related to stellar periodicity because (1) the empirical estimation may be not reliable \citep[e.g.][]{2015A&A...577L...3T,2016Natur.529..181V} and (2) the differences between the results of gyrochronology relation 
and ours are less than 1 order of magnitude. However, the observation mode of TESS is unique, which may limit the stellar period estimation due to some stars are observed for only 27 days. A similar discussion can also be found in the Appendix B of our previous work \citep{2020ApJ...890...46T}. Therefore, we expect that accurate periodic estimation that can be done for TESS targets, like that for Kepler targets \citep{2014ApJS..211...24M}.

\subsection{Active Flare Stars}
\label{sec:Active flar stars}
\begin{figure*}[!h]
	\centering
	\gridline{\fig{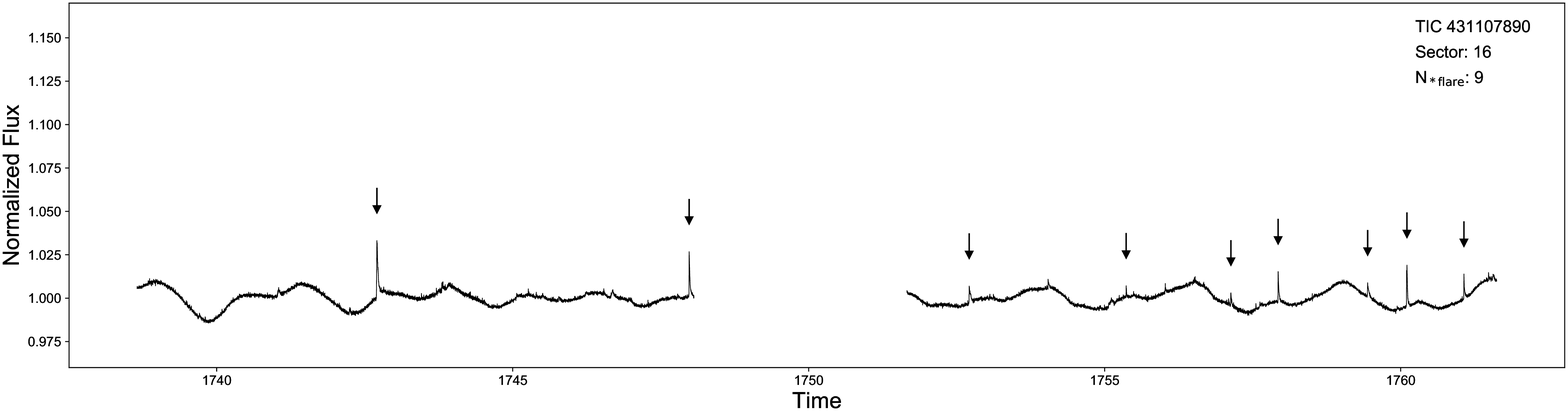}{0.8\textwidth}{}}
	\gridline{\fig{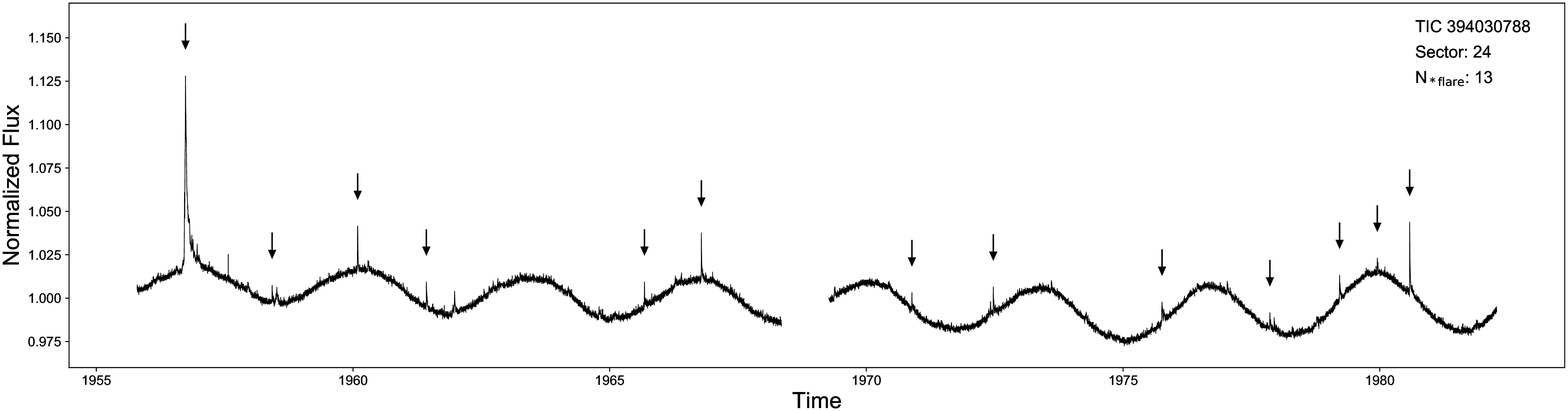}{0.8\textwidth}{}}
	\caption{Light curves of the most active parts of TIC 431107890 and TIC 394030788. Small black arrows mark the superflares of the star. Star TIC 431107890 has the highest flare frequency ($f_{*}$), and TIC 394030788 has the most superflares ($N_{\rm *flare}$). A continuous light curve is cut into two segments with an obvious gap between them, which is caused by data transfer of TESS at each orbit perigee.}
	\label{fig:TIC_431107890}
\end{figure*}

In this work, we also calculate the flare frequency for a single flare star as 
\begin{equation}\label{equ:activefrequency}
	f_{*}=\frac{N_{\text {*flares}}}{\tau_{ {*}}},
\end{equation}
where $N_{\text {*flares}}$ is the count of superflares on a star, and $\tau_{ {*}}$ is the effective observing time length for the star. We sort 311 flare stars according to their flare frequency ($f_{*}$) and number of superflares ($N_{\text {*flares}}$) in Table \ref{tab:flarestars}. For two stars, TIC 431107890 has the highest flare frequency, and TIC 394030788 exhibits the most superflares. The light curves of their most active parts are shown in Figure \ref{fig:TIC_431107890}. Here we use two active stars (TIC 43472154 and TIC 364588501) from \citet{2020ApJ...890...46T} for comparison. Star TIC 431107890 has the highest flare frequency in this work, but it is still less active than TIC 43472154 which generates 16 superflares within just one observing sector. Star TIC 394030788 produces 76 superflares in just eight sectors, and this star is more active than the star TIC 364588501, which generates 63 superflares in 13 sectors. Statistically, we do not find any stars in this work that generate superflares as much as many superflares as TIC 364588501 produced (20 superflares) in the sector 5. However, TIC 394030788 from this work is still worth analyzing in detail, like \citet{2020MNRAS.494.3596D} did for TIC 364588501. 

In this work, the most energetic superflare is the one that releases about $1.24\times 10^{36}$ erg energy from TIC 420137030. However, this is 1 order of magnitude less than the largest superflare with energy $1.77\times 10^{37}$ erg from TIC 93277807 \citep{2020ApJ...890...46T}. So, the star TIC 93277807 is still interesting for exploring its possible superflare mechanisms. Also, we will discuss this topic in Section \ref{sec:sindex and rvar}. We still expect that ground-based follow-up observations and TESS extended missions could provide much more valuable data about this star.

\subsection{Planet Candidates}
\label{sec:Planet}
Stellar activities are significantly relevant to the habitability of planets, and star-planets interactions will improve our understanding of space weather \citep[e.g.][]{2016NatGe...9..452A,2018SSRv..214...21R,2019LNP...955.....L}, and planet atmosphere changes by impact of stellar flares \citep{2020NatAs.tmp..248C}.

It has been argued that close-in Jovian plants may cause an enhancement of stellar magnetic activity. \citet{2000ApJ...533L.151C} suggested the enhancement of stellar activity caused by the tidal and/or magnetic interactions between a host star and a hot Jupiter. \citet{2003ApJ...597.1092S} found that the enhancement of Ca \RNum{2} H and K emission is synchronous with Jupiter's orbit in HD 179949. A similar synchronized Ca \RNum{2} enhancement were reported in other stars, e.g. $\upsilon$ And, HD 189733, HD 73256, and $\tau$ Boo \citep[e.g.][]{2005ApJ...622.1075S,2008ApJ...676..628S}. Theoretically, \citet{2000ApJ...529.1031R} proposed that superflares are caused by magnetic reconnection between fields of a host star and a hot Jupiter. They argued that superflares only happen on those stars hosting close-in giant planets, and our Sun could not exhibit superflares. Through MHD numerical simulations,  \citet{2004ApJ...602L..53I} suggested that flare-like activity with the energy comparable to that of the typical solar flares could be triggered by interaction of the magnetosphere of close-in giant planets with the coronal magnetic field of a host star.

In this work, we use the Exoplanet Follow-up Observing Program for TESS (ExoFOP-TESS\footnote{https://exofop.ipac.caltech.edu/tess/}) 
to cross-match flare stars in our database. Only three targets (0.96\%) of 311 flare stars may host planets, and 204 stars (0.91\%) of 22,228 nonflare stars may host planets. We do not find a significant fraction excess of flare stars hosting planet candidates compared with nonflare stars. Besides, after adding 400 flare stars and 25,334 nonflare stars from TESS first-year observation \citep{2020ApJ...890...46T}, these two fractions are 0.89\% for all 711 flare stars and 0.84\% for all 47,563 nonflare stars, respectively. The above comparisons support that the idea that star-planet interactions may not be the main mechanism for generating superflares of solar-type stars.

We get three planet candidates listed in Table \ref{tab:planets}. From their properties derived by the automatic pipeline of TESS, these two candidates TOI 1254.01 and TOI 1425.01 are up to requirements of hot Jupiter (with orbiting period $P_{\rm orbit} < 10\;{\rm days}$, and planetary radius $8.0\;{R_{\oplus}}\leqslant R_{\rm	planet}  \leqslant 32.0\; {R_{\oplus}}$ \citep{2012ApJS..201...15H}). Hot Jupiter was believed to be an important factor of generating superflares \citep[e.g.][]{2000ApJ...529.1031R,2004ApJ...602L..53I}, but recent works show this mechanism is rare for superflares on solar-type stars \citep[e.g.][]{2012Natur.485..478M,2013PASJ...65...49S,2020ApJ...890...46T}. Besides, TOI 1429.01 is an ultrashort period (UPS) planet candidate, as its orbit period less than 1 day \citep{2018NewAR..83...37W}. A UPS planet is close to its hosting star, so star-planet interaction may evaporate its atmosphere. We only refer to those planetary properties from ExoFOP-TESS, which may be not accurate for further specific analysis, but these three planet candidates are worthy of being observed by further confirmations.

\subsection{Superflare Energy Versus Stellar Period}
\label{sec:energy-period}
\citet{2019ApJ...876...58N} used the plot of superflare energy and stellar period and found a decreasing trend between the upper limit of energy and stellar periodicity. This discovery was based on the catalog from \citet{2014ApJS..211...24M}, which applied accurate period estimations for Kepler targets. 
In our previous work \citep{2020ApJ...890...46T}, we found that only the tail part (period more than a few days) of the plot showed this trend, as the TESS unique observation mode will limit the estimation of stellar periods. 

\begin{figure*}[!h]
	\centering
	\gridline{\fig{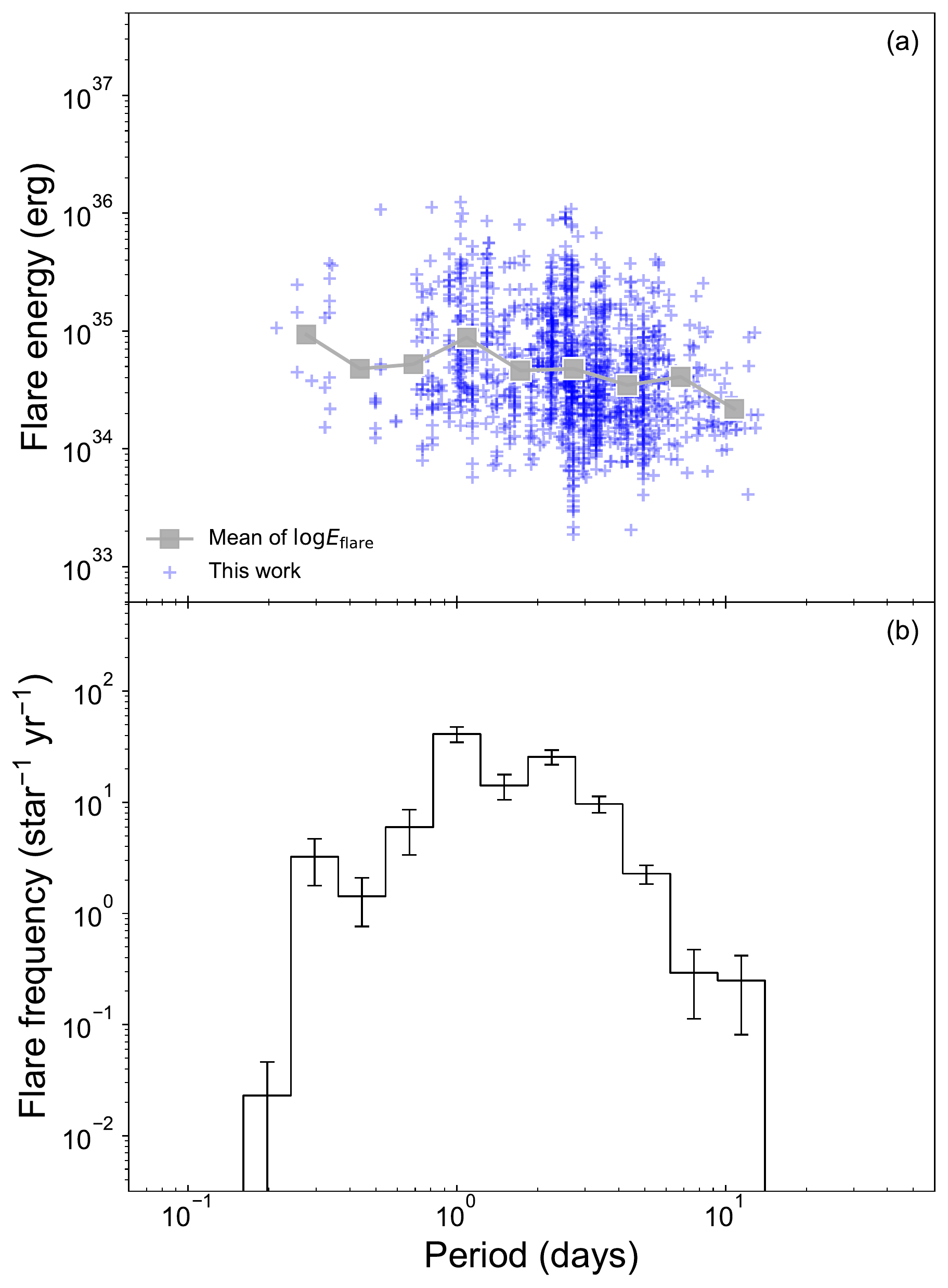}{0.49\textwidth}{}
		\fig{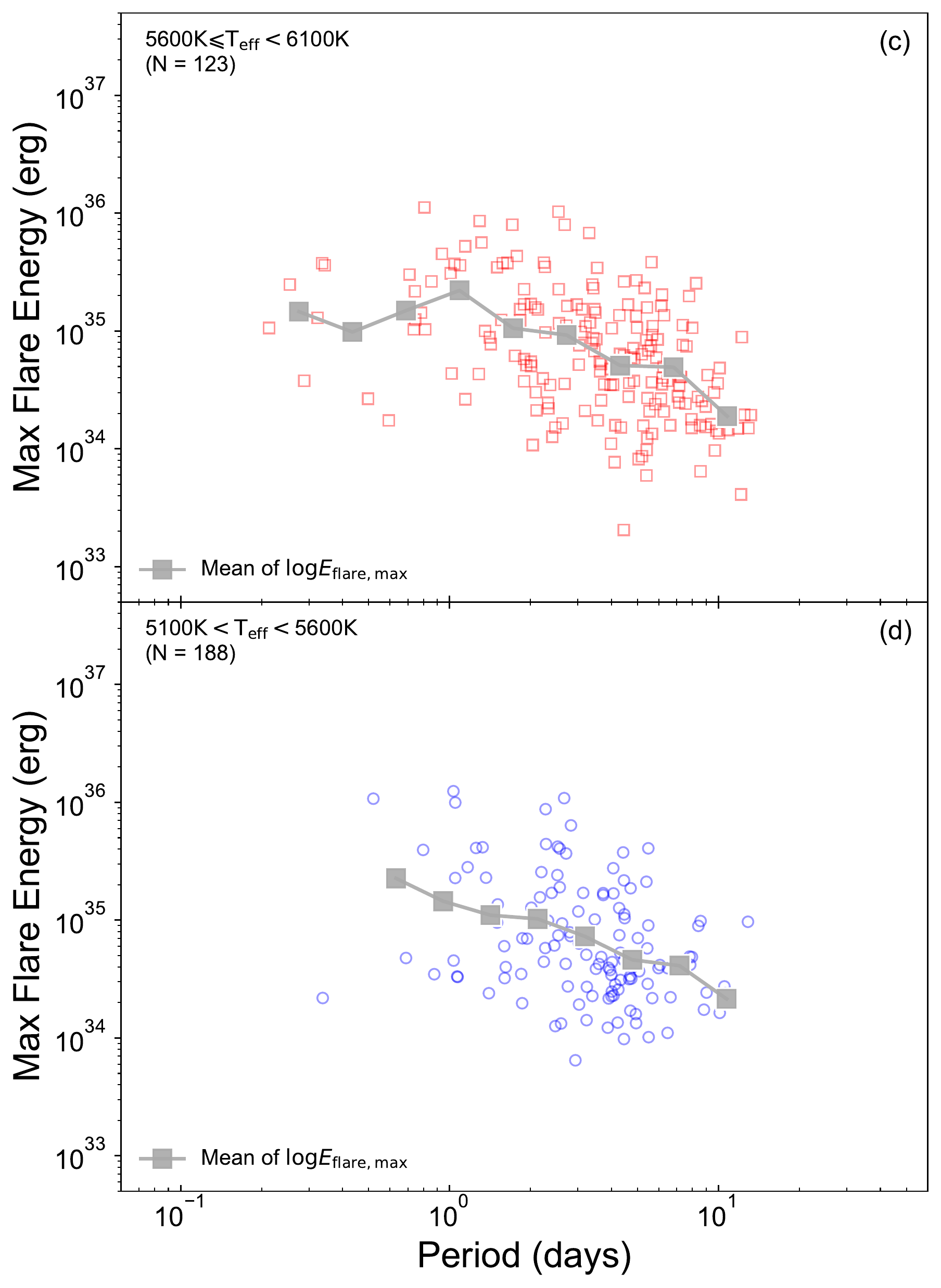}{0.49\textwidth}{}
	}
	\caption{(a) Plot of stellar period versus superflare energy. Gray squares indicate the mean values of $\log E_{\rm flare}$ in nine bins of stellar period.
		(b) Flare frequency distribution as a function of stellar period. Errors are calculated from the square root of counts in each period bin.
		Panels (c) and (d) are plots of maximum flare energy versus stellar period. Gray squares are same as those in  panel (a).}
	\label{fig:period energy}
\end{figure*}

In this work, from Figure \ref{fig:period energy}(a), the descending tendency is only apparent for stars with periods of more than a few days (tail part). We find no strong decreasing trend with the mean of $\log E_{\rm flare}$ in nine period bins. Besides, the flare frequency is decreasing with periods over a few days, as shown in panel (b). Because the upper energy limit in each period bin from the work of \citet{2019ApJ...876...58N} is just the maximum flare energy for the star, we use the maximum energy on the star instead of attaching all superflare energies. We also separate stars according to their surface temperatures, as described in the legends of Figures \ref{fig:period energy} (c) and (d). These two panels both show an upper-limit decreasing trend at the tail parts, as we see in panel (a). Gray squares indicate for the mean value of $\log E_{\rm flare}$ in each period bin, which shows a more apparent descending tendency. 

\subsection{Occurrence Frequency Distribution}
\label{sec:frequencies}
\begin{figure*}[!h]
	\centering
	\includegraphics[width=1\linewidth]{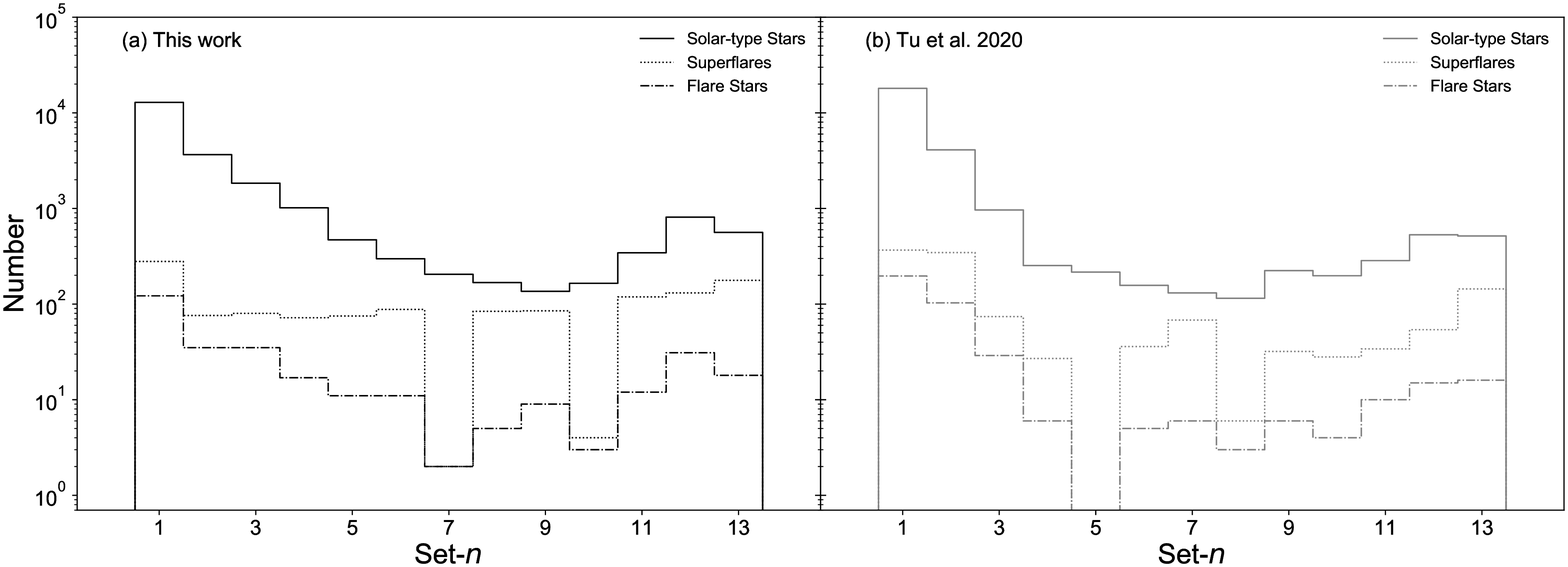}
	\caption{Distributions of stars and superflares in each Set-$n$. In panel (a), the solid, dotted, and dashed-dotted lines represent the number of solar-type stars, superflares, and flare stars, respectively. The corresponding results are listed in Table \ref{tab:SetN}. In panel (b), the results from the TESS first-year observation \citep{2020ApJ...890...46T} are shown for comparison.}
	\label{fig:SetN}
\end{figure*}

Because TESS changes its observing field every $\sim 27$ days, the targets observed by TESS have different durations of observing time. We still use the same functions as we specifically described in our previous work, which is based on the frequency calculation of \citet{2012Natur.485..478M}. We also subdivide those targets in our database into different subsets called as Set-$n$, where $n$ equals 1 - 13 and stands for the umber of sectors in which the targets have been observed. The distribution of Set-$n$ is shown in Figure \ref{fig:SetN}. Specifically, counts of superflares, flare stars, and solar-type stars are shown in Table \ref{tab:SetN} according to their surface temperatures and periods. In Figure \ref{fig:SetN}, we also show the Set-$n$ distribution in our previous work \citep{2020ApJ...890...46T} as panel (b) for comparison. It seems that the Set-$n$ distributions are similar. 

We use the following function to calculate flare frequencies \citep{2020ApJ...890...46T}.
\begin{equation}\label{equ:frequency}
	f_{n}=\frac{N_{\text {flares}}}{N_{\text {stars}} \cdot \tau_{ {n}}
		\cdot \Delta E_{\text {flare}}},
\end{equation}
where the subscript $n$ is same as the $n$ of Set-$n$. $N_{\rm flares}$ and $N_{\rm stars}$ represent counts of superflares and solar-type stars in the same Set-$n$. $\Delta E_{\text {flare}}$ gives the energy range in each bin, and 
\begin{equation}\label{equ:tau_n}
	\tau_{ {n}} = n \times 22.64 \; \text{days},
\end{equation}
where 22.64 represents the mean of the effective observing time of one sector in the TESS second-year observation, as TESS does not fully observe in 27 days for one sector. The flare frequency is derived as 
\begin{equation}\label{equ:13frequency}
	f=\frac{{\sum}f_{n}}{13},
\end{equation}
which represents the mean frequency of 13 Set-$n$.

\begin{figure*}[!h]
	\gridline{\fig{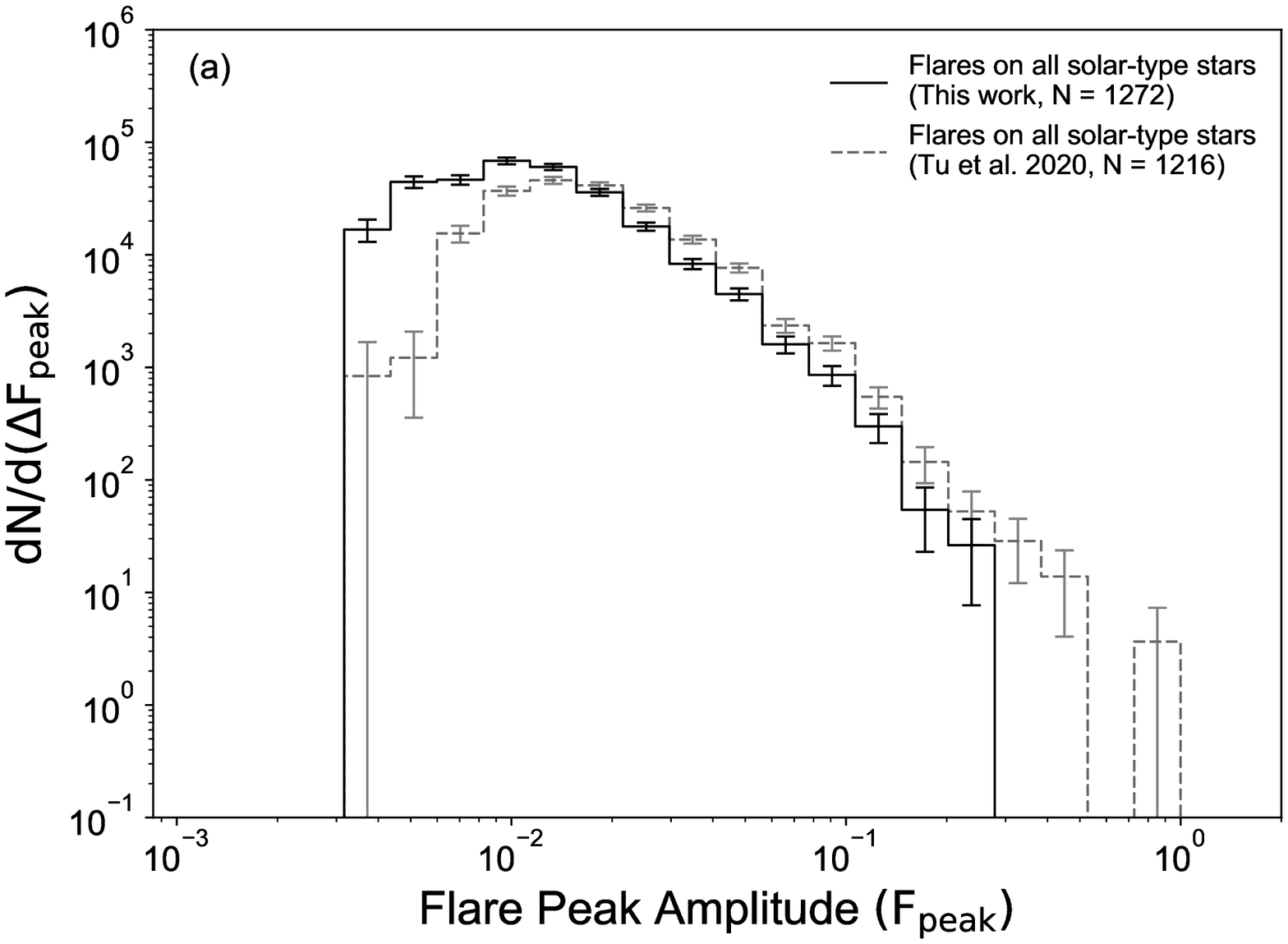}{0.5\textwidth}{}
		\fig{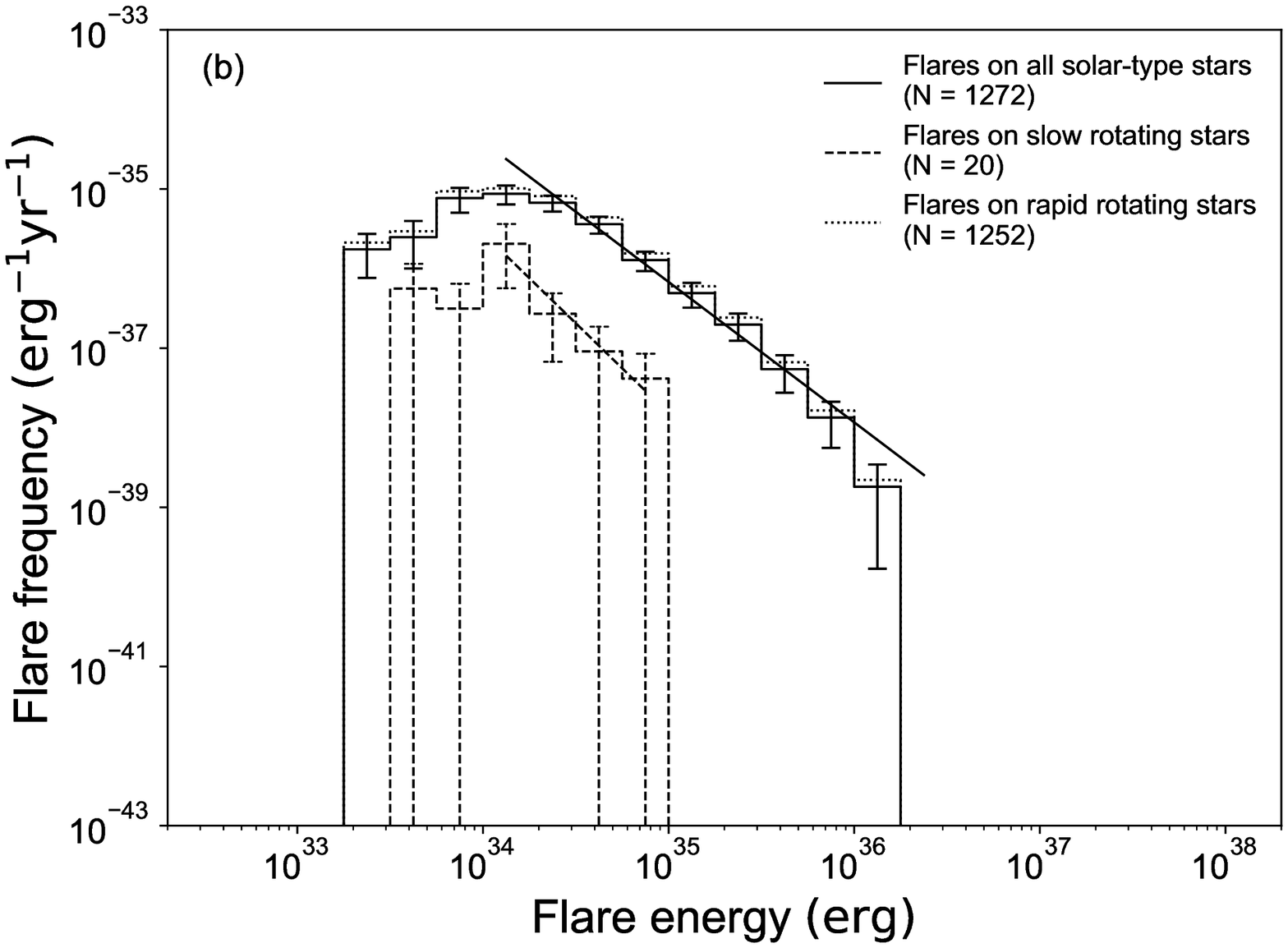}{0.5\textwidth}{}
	}
	\gridline{\fig{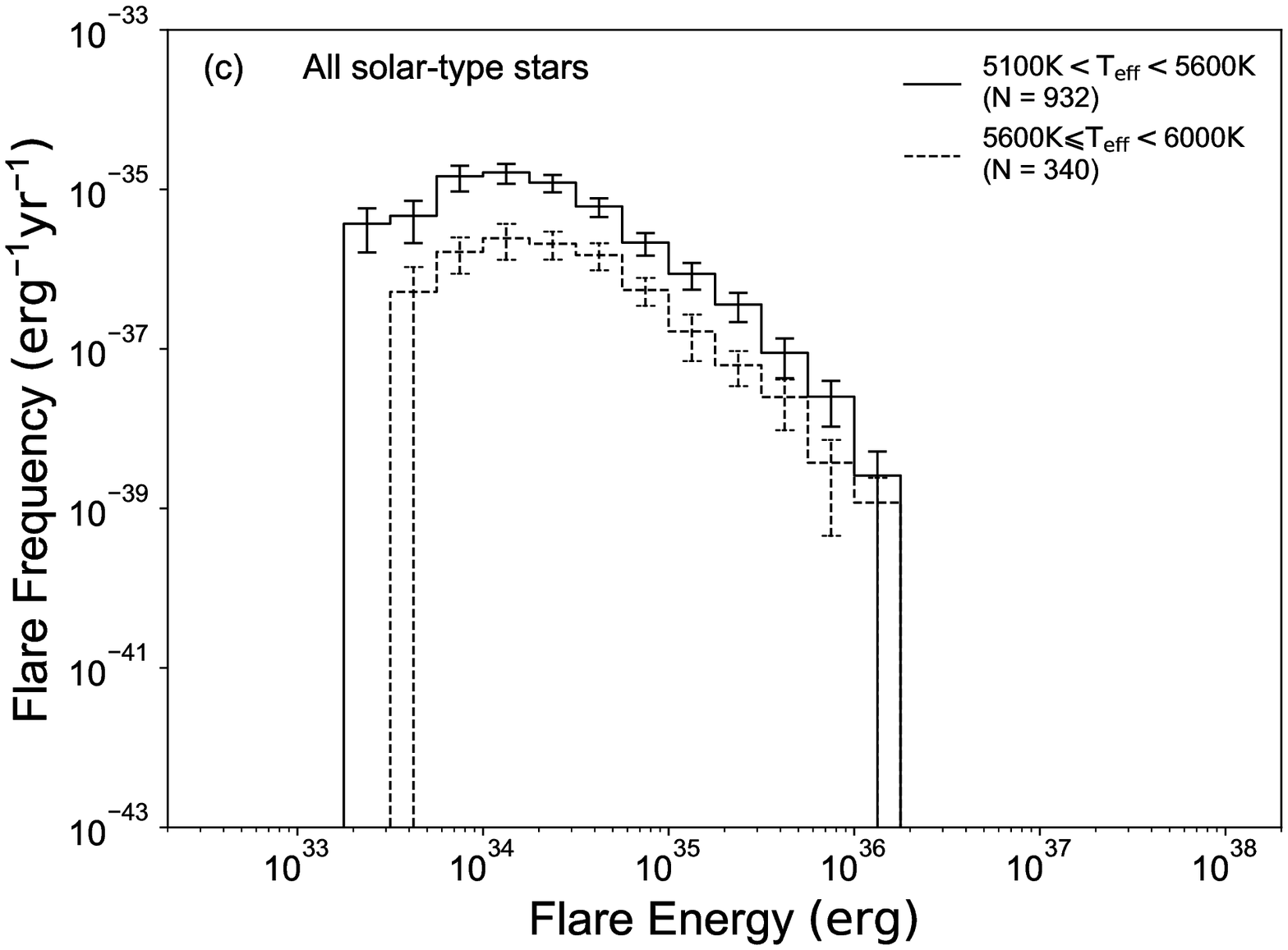}{0.5\textwidth}{}
		\fig{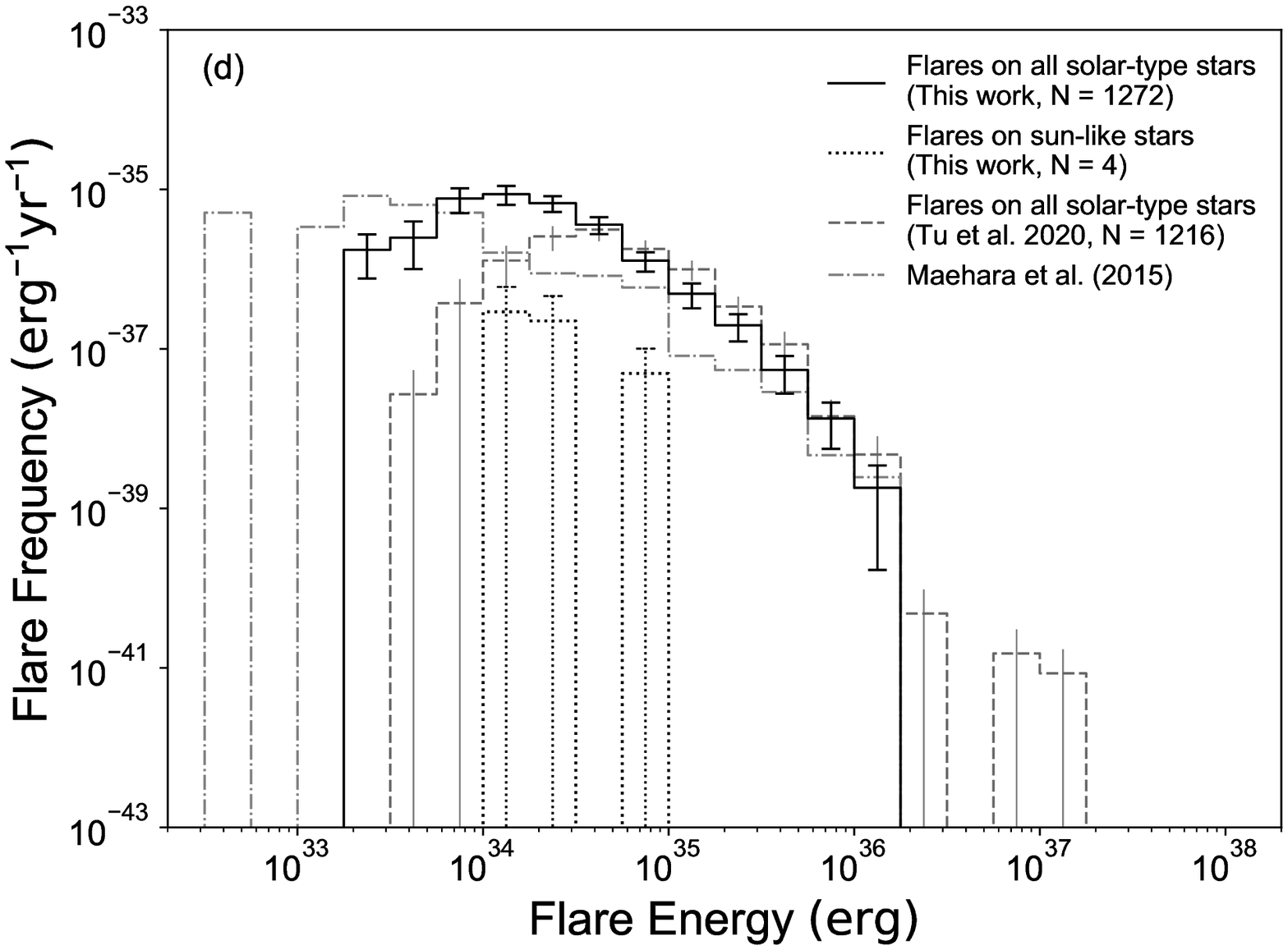}{0.5\textwidth}{}
	}
	
	\caption{Flare frequency as a function of flare peak amplitude ($F_{\rm peak}$) and flare energy. Error bars are calculated by the square root of the flare counts in each bin.
		(a) Flare frequencies as a function of $F_{\rm peak}$ for all 1272 superflares. For comparison, those results of \citet{2020ApJ...890...46T} are shown by gray dashed step lines.
		(b) Flare frequency as a function of flare energy, where solid, dashed, and dotted outlines represent superflares on all flare stars and slowly and rapidly rotating stars, respectively. The straight solid line represents the power-law correlation of $d N / d E \propto E^{\gamma}$ with index $\gamma \sim -1.76 \pm 0.11$. The straight dashed line gives power-law index $\gamma \sim -2.30 \pm 0.66$.
		(c) Flare frequencies for stars with different stellar surface temperatures are shown.	
		(d) Same as panel (b) but including frequencies from other data sets. All 1272 superflares are represented by solid step lines, which are the same as panel (b). Dotted outlines represent frequencies of 4 superflares on Sun-like stars, which are defined in Section \ref{sec:solar-type data}. For comparison, 1216 superflares of \citet{2020ApJ...890...46T} and 187 superflares of \citet{2015EP&S...67...59M} are represented in gray dashed and dashed-doted step lines, respectively.  
		\label{fig:oc_fre}}
\end{figure*}

In Figure \ref{fig:oc_fre}(a), we compare the flare peak amplitude distribution of these two-year observations. Here $F_{\rm peak}$ is $F_{\rm flare}$ of Equation (\ref{equ:flare_flux}) at peak times. The solid line is higher than the gray dashed line, with peak amplitudes less than $10^{-2}$. Contrarily, the solid line is slightly lower than the gray dashed line, with peak amplitudes greater than $10^{-2}$. 
The results indicate that the detected superflares in this work generally have a lower $F_{\rm peak}$ than those of \citet{2020ApJ...890...46T}. As peak amplitude is positively correlated with flare energy, if more superflares have lower peak amplitudes, it will pull down the overall energy distribution, just like what is shown in Figure \ref{fig:Mflare}(d). Meanwhile, the superflares with $F_{\rm peak} \gtrsim 10^{-2}$ are lower than those of previous work, which may even reduce the overall energy of their distribution. From the view of peak amplitude, the superflares in this work are overall weak.

In Figures \ref{fig:oc_fre} (b) and (c), we get similar results as in previous literature \citep[e.g.][]{2012Natur.485..478M,2013ApJS..209....5S,2015EP&S...67...59M}; (1) fast-rotating stars have higher superflare occurrence frequencies than slowly rotating ones, as shown in panel (b), and (2) cooler stars generate superflares more frequently than hotter stars, as shown in panel (c). Besides, in Figure \ref{fig:oc_fre}(b), dotted line is basically overlapped with the solid line. This indicates that the solar-type stars observed by TESS are dominated by rapidly rotating stars. In this work, 19,101 stars have periods less than 10 days, which is 85\% fraction of all solar-type stars. This same number of this fraction is also deduced from TESS first-year observations \citep{2020ApJ...890...46T}. In Figure \ref{fig:oc_fre}(d), solid black lines are above dashed-dotted gray lines. This may also be explained by the large fraction of rapidly rotating stars in our database. In this work, 85\% of the stars have a period of less than 10 days, and this number is 32\% in \citet{2015EP&S...67...59M}. That stellar periodicity is related to their age, so young stars rotate faster than old stars \citep{1972ApJ...171..565S,2003ApJ...586..464B}. Furthermore, young stars are more active and have higher superflare frequencies than old stars. 

The flare frequencies of Sun-like stars are only deduced from four superflares in panel (d), so we may not be able to statistically get solid conclusions from that.

\begin{figure*}[!h]
	\centering
	\gridline{\fig{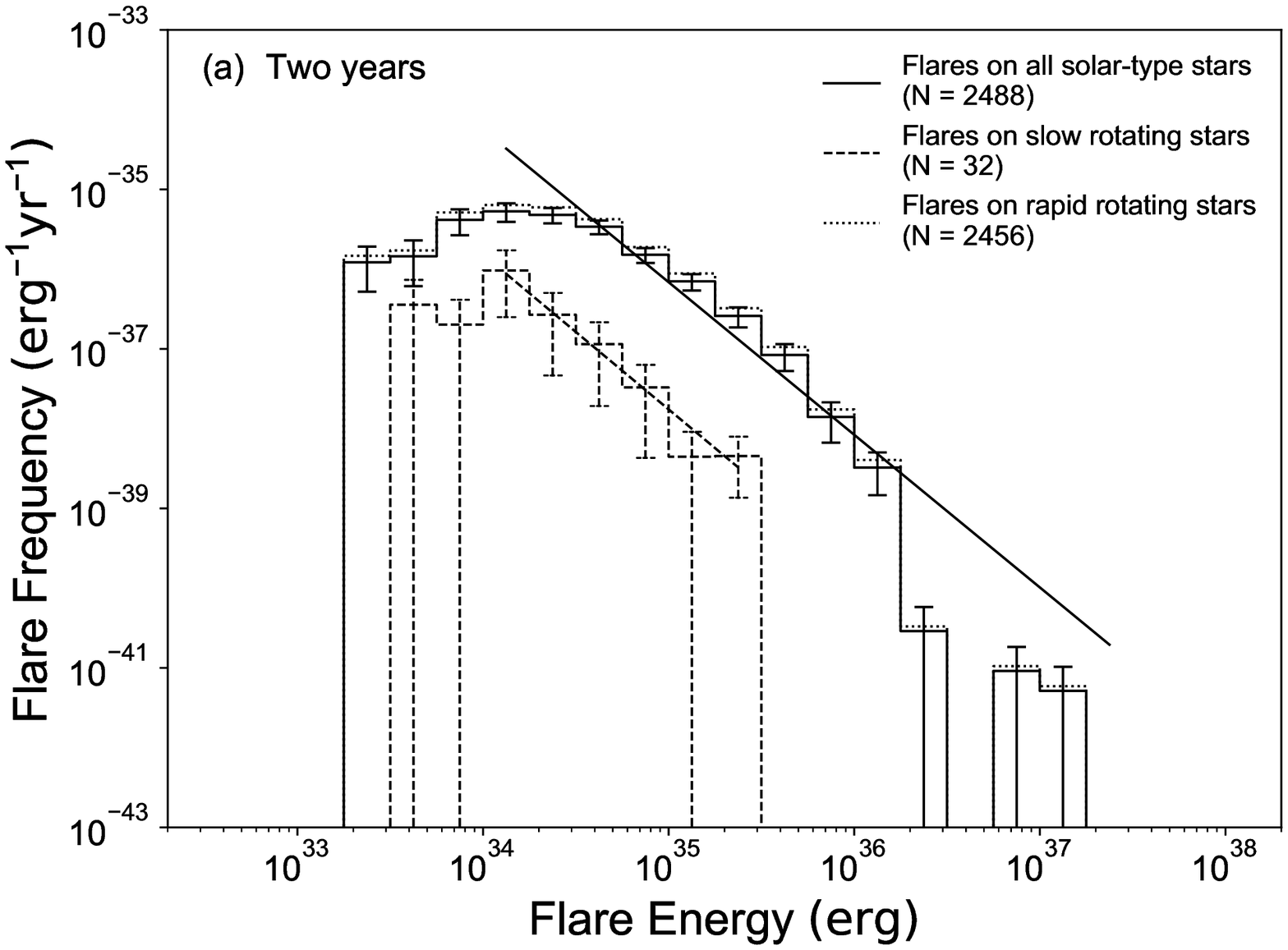}{0.5\textwidth}{}
		\fig{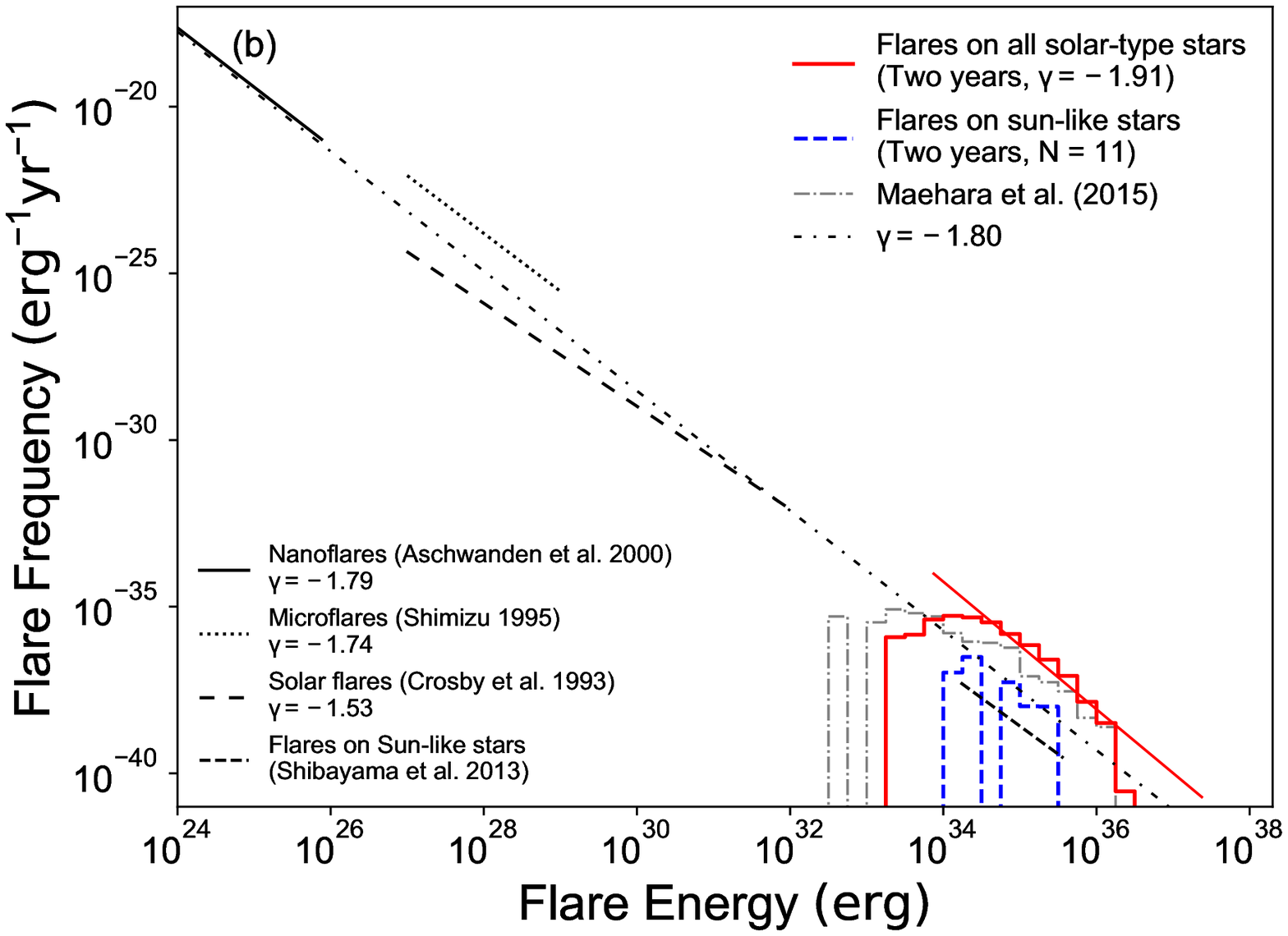}{0.5\textwidth}{}
	}
	\caption{(a) Flare frequency of all superflares from TESS 2 yr observations. The power-law index of $d N / d E \propto E^{\gamma}$ is $\gamma = -1.91 \pm 0.07$ for all superflares, which is shown by the solid line. Slowly rotating stars give $\gamma = -1.94 \pm 0.31$, which is represented by the dashed line. (b) Solar flare frequencies, including solar nanoflares \citep{2000ApJ...535.1047A}, microflares \citep{1995PASJ...47..251S}, and flares \citep{1993SoPh..143..275C}, are also shown. Each corresponding power-law index $\gamma$ is listed in the legend. We also import the results of Sun-like stars from Kepler observations \citep{2013ApJS..209....5S} as the black densely dashed  line. The black dashed-dotted line represents a power-law correlation with index $\gamma=-1.8$. Results from \citet{2015EP&S...67...59M} are shown by histograms with gray dashed-dotted outlines. We import superflares from TESS 2 yr observations, which are represented by the red histogram. Eleven superflares are from Sun-like stars and represented by blue dashed lines.}
	\label{fig:compare flare}
\end{figure*}

Flare frequency distribution as a function of energy can be applied by a power-law relation as 
\begin{equation}
	\frac{d N }{d E}\propto E^{\gamma}.
\end{equation}
Through linear regression, we get a power-law index $\gamma \sim -1.76 \pm 0.11$ for all superflares, and $\gamma \sim -2.30 \pm 0.66$ for superflares on slowly rotating stars. For all superflares, $\gamma \sim -1.76 \pm 0.11$ is lower than that of the TESS first-year observation ($\gamma \sim -2.16 \pm 0.10$) and \citet[][$\gamma \sim 2.2$]{2013ApJS..209....5S}. We also combine the superflare database from our previous work with data from this work and get $\gamma \sim -1.91 \pm 0.07$ for all superflares, as shown in Figure \ref{fig:compare flare}(a). For superflares on slowly rotating stars, $\gamma \sim -1.94 \pm 0.31$. One thing needs to be addressed: that combining 2 yr superflares may not be a strict behavior, as the TESS pipeline and accuracy of our visual inspection are all improved (refer to Section \ref{sec:solar-type data}). However, we do not apply the new method and TESS newly released data for reprocessing TESS first-year observation, as they may not significantly affect the robustness of conclusions in this work. Anyway, we can still roughly get overall flare frequencies as a function of energies from TESS 2 yr observations. 

For comparison with solar flare frequencies, we use different kinds of solar flares from some databases and show them in Figure \ref{fig:compare flare}(b). It is obvious and fascinating that, as shown in Figure \ref{fig:compare flare}(b), solar nanoflares \citep{2000ApJ...535.1047A}, solar microflares \citep{1993SoPh..143..275C}, solar flares \citep{1995PASJ...47..251S}, and superflares on solar-type stars from {\em Kepler } \citep{2013ApJS..209....5S, 2015EP&S...67...59M}, and from TESS \citep[][and this work]{2020ApJ...890...46T} basically distribute around $d N / d E \propto E^{-1.8}$ (black dashed-dotted line). This may indicate same mechanism is shared by these different types of flares. As the flare energy increases, their frequency decreases. It is obvious that TESS gets overall higher superflare frequencies than Kepler after we use the combined superflare database from the TESS 2 yr observations. The reason why superflare frequencies from this work (red solid lines) are higher than those of \citet[][gray dash-dotted lines]{2015EP&S...67...59M} is addressed in the above paragraph. In total, the superflare frequency for Sun-like stars is apparently higher than that of \citet{2013ApJS..209....5S}. However, only 11 superflares are observed by TESS 2 yr missions. We expect that more superflares on Sun-like stars can be detected by TESS extended missions, and other practicable methods can more accurately estimate stellar periods for TESS targets, which will strengthen the reliability of selecting superflares on Sun-like stars. Interestingly, similar power-law distributions are also found in different astronomical phenomena, i.e.,  gamma-ray bursts \citep{Wang2013,Yi2016,Lyu2020}, fast radio bursts \citep{Wang2017,Cheng2020}, and black holes \citep{Li2015,Wang2015,Yan2018,Yuan2018}.

\subsection{Superflare Energy and Duration}
\label{sec:e-d-relation}
\begin{figure*}[!h]
	\centering
	\gridline{\fig{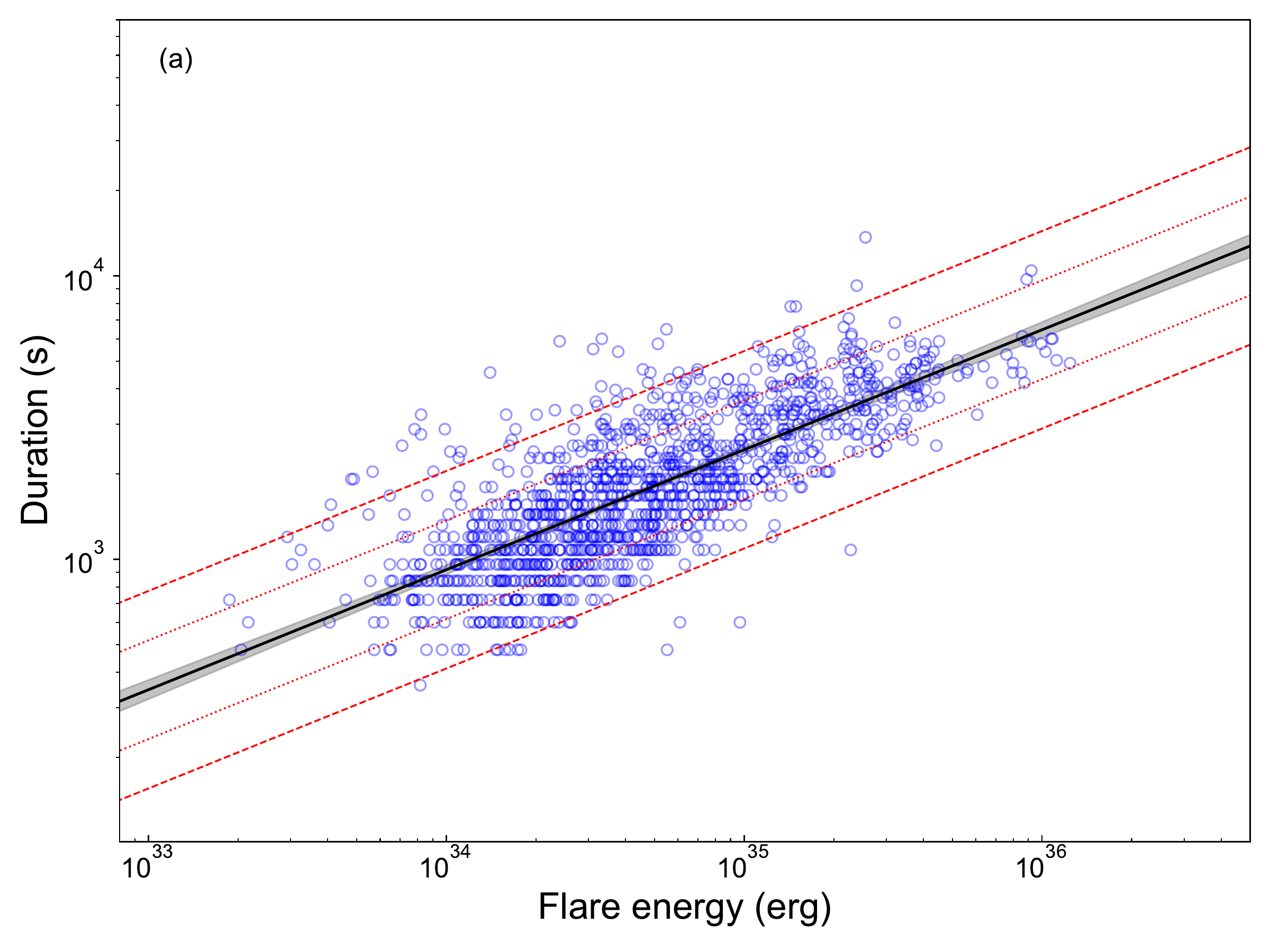}{0.49\textwidth}{}
		\fig{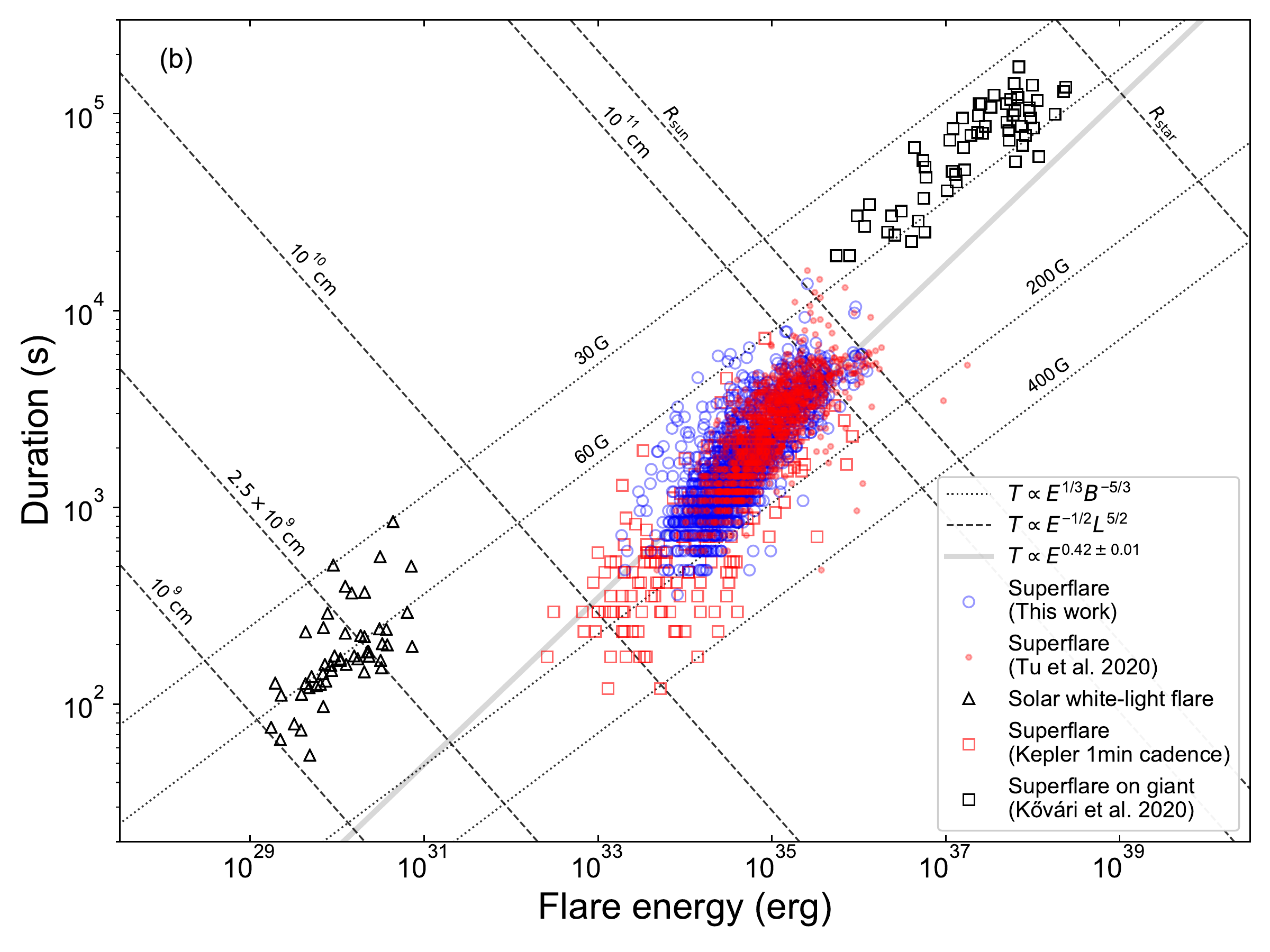}{0.49\textwidth}{}
	}
	\caption{Plots of superflare energy versus duration in log scale. (a) The black solid line represents the best-fitting results of the power-law correlation $T_{\text {duration }} \propto E^{0.42\pm 0.01}$. The 95\% confidence interval of the uncertainties is shown in gray shading. The $1\sigma$ and $2\sigma$ intervals of extra variability (same as $\sigma_{v}$ in \citet{2018ApJ...869L..23T}) are shown as red dotted and dashed lines, respectively. (b) Blue circles, red dots, and red squares represent superflares from this work, \citet{2020ApJ...890...46T}, and \citet{2015EP&S...67...59M}, respectively. Black triangles indicate for solar white-light flares \citep{2017ApJ...851...91N}. Black squares represent superflares on the late-type giant star KIC 2852961 \citep{2020A&A...641A..83K}. We also use a gray line to represent the correlation line of $T_{\text {duration }} \propto E^{0.42\pm 0.01}$ in panel (b).
	The power-law correlations of Equations (\ref{equ:T-E-L}) and (\ref{equ:T-E-B}) are shown as dotted and dashed lines, respectively. The magnetic field strength ($B$) and flaring length scale ($L$) have been set to different values \citep{2017ApJ...851...91N}. Note that $R_{\rm sun}$ and $R_{\rm star}$ are the diameter of the Sun \citep{2017ApJ...851...91N} and KIC 2852961 \citep{2020A&A...641A..83K}, respectively.
	}   	\label{fig:dura-energy}
\end{figure*}

Statistically, we find that superflares and solar flares may have similar physical mechanisms, as discussed in Section \ref{sec:frequencies}. Here we also apply power-law correlation fitting between the flare energy and duration. In Figure \ref{fig:dura-energy}(a), we get the power-law correlation  \begin{equation}\label{equ:energyduration}
	T_{\text {duration }} \propto E^{\beta}, 
\end{equation}
with $\beta = 0.42 \pm 0.01$. Here $\beta = 1/3$ is derived from the magnetic reconnection theory \citep{2015EP&S...67...59M}, which is successfully applied to solar flares. Here $\beta = 0.42 \pm 0.01$ is the same as that of our previous work \citep{2020ApJ...890...46T}. And \citet{2015EP&S...67...59M} got $\beta = 0.39 \pm 0.03$. For solar flares, \citet{2018ApJ...869L..23T} got $\beta = 0.33 \pm 0.001$. So, superflares generally get a higher $\beta$ value than one=third. 
This reinforces our conclusion that solar flares and superflares may have similar mechanisms not only from their occurrence frequency distributions but also from the close correlation between superflare energies and durations. However, the obviously higher value of $\beta$ may also point out that there must be some differences between mechanisms of solar white flares and superflares on solar-type stars. 

If we consider magnetic strength as a variable, the Equation \ref{equ:energyduration} will be reformed as
\begin{equation}\label{equ:T-E-B}
	T_{\rm {duration }} \propto E^{1/3}B^{-5/3},
\end{equation}
\begin{equation}\label{equ:T-E-L}
	T_{\rm {duration }} \propto E^{-1/2}L^{5/2},
\end{equation}
where $B$ and $L$ are the coronal magnetic field strength and flare length scale, respectively. \citet{2017ApJ...851...91N} derived these two relations in detail, and we also use the coefficients of $B$ and $L$ also from their work to draw scale lines in Figure \ref{fig:dura-energy}(b). Solar white-light flares of \citet{2017ApJ...851...91N}, superflares from \citet{2015EP&S...67...59M}, and our previous work are all imported into this figure. Compared with the red dots in the figure, superflares from this work (blue circles) have a larger overlapping area with those (red squares) from \citet{2015EP&S...67...59M}. Even if the TESS filter is less sensitive to weak superflares, as we have discussed in Section \ref{sec:energy distributions}, there are still some less energetic superflares with $E_{\rm flare} \lesssim 10^{34}$ erg, which we have selected through the improved visual inspection. Overall, superflares have a higher coronal magnetic field strength and larger length scale than those of solar white-light flares. 

Recently, statistical research on superflares on giant stars has been done by \citet{2020A&A...641A..83K}. They used Kepler and a small part of the TESS data and searched 60 superflares on the star KIC 2852961. It is also interesting to compare superflares on solar-type stars with those on giant star KIC 2852961. These superflares are shown by black squares in Figure \ref{fig:dura-energy}(b). They seem to be an extension part of the superflares on solar-type stars with greater flare energies and longer durations. From the scaling line of the coronal magnetic field strength ($B$) and flaring length scale ($L$), we find that the superflares on KIC 2852961 basically have a coronal magnetic strength between 30 G and 60 G, which is smaller than that of superflares on solar-type stars. From the view of flaring length scale, superflares on solar-type stars are basically under the limitation of solar diameter ($R_{\rm sun}$). However, superflares on giant stars basically have a much larger $L$ than solar-type stars. The radius of this giant star is reported as $R_{*} = 13.1{R_ {\odot}}$. According to Figure \ref{fig:dura-energy}(b), the flaring lengths of superflares on KIC 2852961 are less than the diameter of this star ($R_{\rm star} = 2R_{*}$). These results also strengthen the conclusion from \citet{2020A&A...641A..83K} that the largest flares of KIC 2852961 have flare loop sizes on the order of $R_{*}$.

\subsection{$S$-index and $R_{\rm var}$}
\label{sec:sindex and rvar}
\begin{figure*}[!h]
	\centering
	\includegraphics[width=0.5\linewidth]{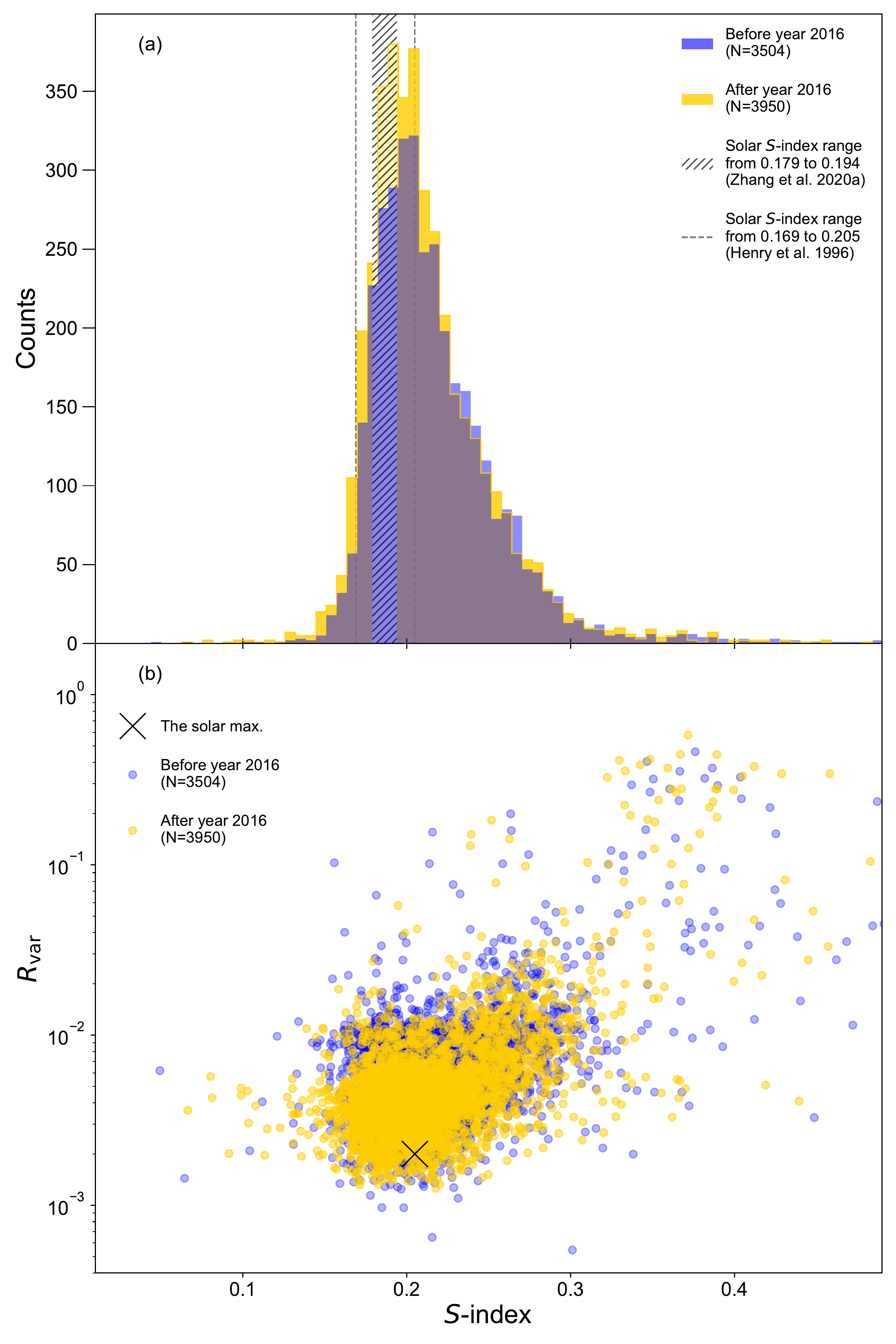}
	\caption{(a) The blue and yellow histograms represent count distributions from LAMOST observations before 2016, and after 2016, respectively. The dashed lines represent the solar $S$-index range according to \citet{1996AJ....111..439H}, who collected $S$-indexes at solar maximum and minimum solar cycles. The gray hatched area masks another range according to \citet{2020ApJ...894L..11Z}, who calibrated the solar $S$-index in the scale of LAMOST. (b) Scatter plot of $S$-index versus $R_{\rm var}$. The black cross represents maximum $S$-index and $R_{\rm var}$ of the Sun, which are equal to $S=0.205$ \citep{1996AJ....111..439H} and $R_{\rm var}=0.0020$ \citep{2020Sci...368..518R}.} 
	\label{fig:allSindex}
\end{figure*}

In Section \ref{sec:rvar} and \ref{sec:sindex}, we have introduced the$S$-index and $R_{\rm var}$, including their calculation methods. In this work, we have 7454 solar-type stars matched with spectra obtained by LAMOST. According to their observing time (before or after 2016), we classify these stars into two subsets, as described in the legend of Figure \ref{fig:allSindex}(a). From the histograms in Figure \ref{fig:allSindex}(a), they do not show obvious differences. The Kolmogorov–Smirnov test also shows that their similarity is over the 99\% confidence level. The result indicates that the observation of LAMOST is more and more stable over many years, and we could make full use of LAMOST to push our understanding of our topic addressed in this work. 
Besides, the $S$-index represents the chromospheric activity of a star, which may not be significantly changed within a couple of years. This idea was proved by \citet{2020ApJ...894L..11Z}. They used 577 stars, of which each star matched more than one record of LAMOST data. After analyzing those spectra, they did not find significant variations of the $S$-index of a star within a couple of years. For the above reasons, in this work, we choose to still use LAMOST data from late 2011 to early 2020, even though only 228 (3\%) of 7454 spectra are observed by LAMOST after 2019 July, when TESS began its second-year mission.

In Figure \ref{fig:allSindex}(a), two ranges of solar $S$-index  are imported. The shaded area represents ranges imported from \citet{2020ApJ...894L..11Z}, who calibrated the solar $S$-index in the scale of LAMOST. The dashed vertical lines are from \citet{1996AJ....111..439H}, who collected $S$-indexes at solar maximum and minimum solar cycles. From the distribution in Figure \ref{fig:allSindex}(a), when we use the upper limit of the $S$-index range, over 52.58\% of the stars have larger $S$-indexes than that of the Sun. 

\begin{figure*}[!h]
	\centering
	\includegraphics[width=0.8\linewidth]{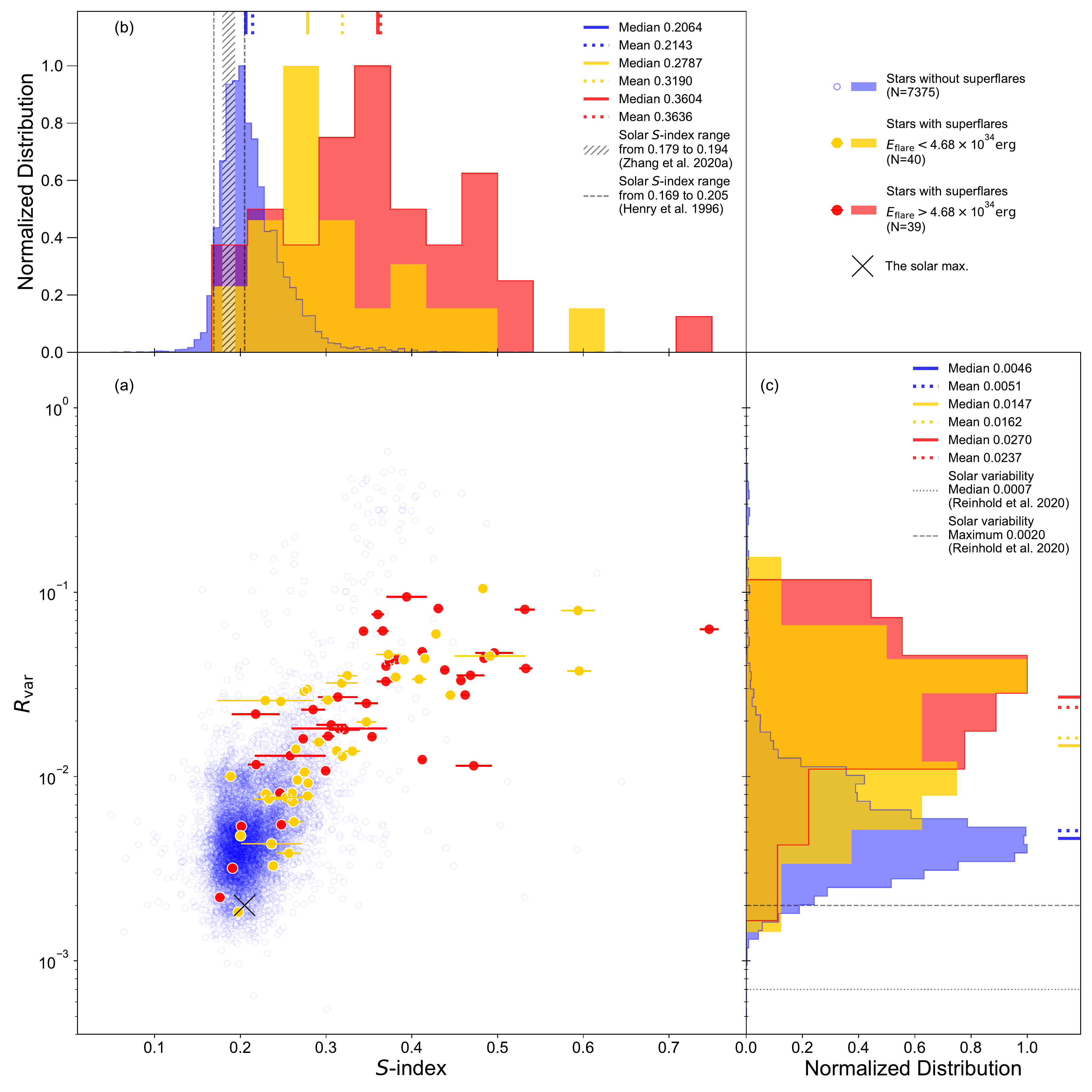}
	\caption{Plots of $S$-index versus $R_{\rm var}$ and their distributions. We set the division of superflare energy at $4.68\times 10^{34}$ erg, which equals a mean value of $\mu = 34.67$ of normal distribution fitting for superflare energy in log scale, and it is also listed in the legend of Figure \ref{fig:Mflare}(a). For the 79 stars with superflares, flare energies over this division are shown in red, and those below this division are shown in yellow. Blue represents 7375 nonflare stars, which do not have any superflares at all. The mean and median values of each distribution are shown by short dotted and solid lines, for which the results are listed in the legends and shown by the same color as the corresponding distribution. The black cross in panel (a) represents the maximum $S$-index and $R_{\rm var}$ of the Sun, which is the same as that in panel (b) of Figure \ref{fig:allSindex}. (b) For the solar $S$-index, we use the same patches and lines shown in Figure \ref{fig:allSindex}. (c) Dotted and dashed lines represent median and maximum values of solar $R_{\rm var}$ from the results of \citet{2020Sci...368..518R}.}
	\label{fig:flare-sindex}
\end{figure*}

\begin{figure*}[!h]
	\centering
	\gridline{\fig{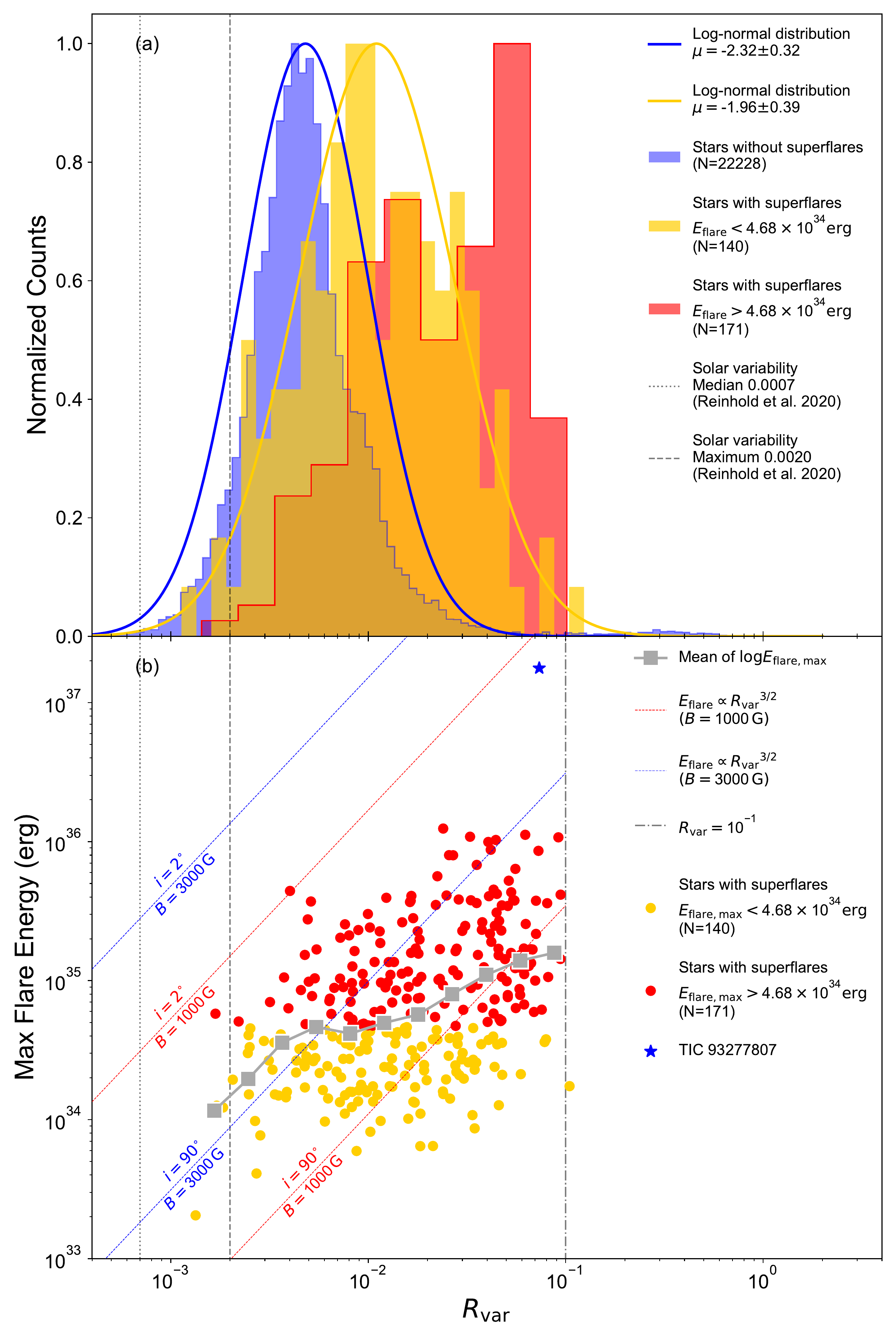}{0.49\textwidth}{}
		\fig{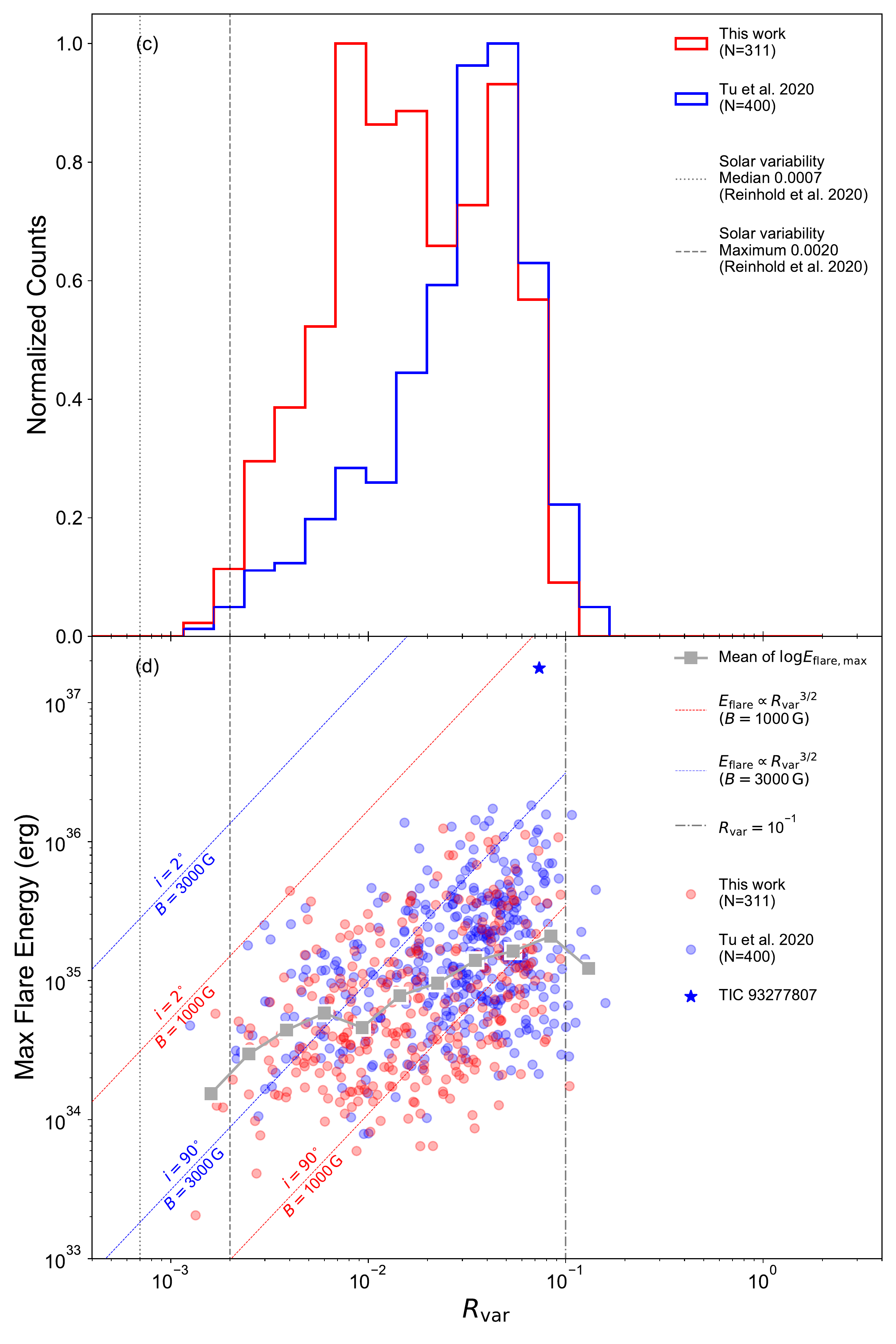}{0.49\textwidth}{}
	}

	\caption{Plots of stellar $R_{\rm var}$ distributions and scatters of $R_{\rm var}$ vs. the maximum energy of the superflare. In these four panels, the solar $R_{\rm var}$ ranges are the same as in Figure \ref{fig:flare-sindex}. (a) Red and yellow represent stars separated by the energy division ($4.68\times 10^{34}$ erg), same as in Figure \ref{fig:flare-sindex}. Blue and yellow histograms are also fitted by the normal distribution function, for which the corresponding results are listed in the legend. In panel (a), we do not apply the same fitting for the red histogram, as it apparently does not show normal distribution. (b) We calculate the mean value of the maximum flare energy in log scale of 11 $R_{\rm var}$ bins, which are marked as gray squares. The blue star represents TIC 93277807. The vertical dashed-dotted line represents $R_{\rm var} = 10^{-1}$. Panels (c) and (b) show similar plots as panels (a) and (b) but use combined data sets from \citet[][400 superflare stars, shown in blue]{2020ApJ...890...46T} and this work (311 superflare stars, shown in red). The red and blue diagonal dashed lines in panels (b) and (d) represent the correlation $E_{\rm flare}\propto R_{\rm var}^{3/2}$ with magnetic field strength $B=1000$ and $3000\,{\rm G}$, respectively. Inclination angles $i = 2^{\circ}$ and $90^{\circ}$ are also included and represented by different lines with labels. This correlation is specifically deduced in Section \ref{sec:energy and rvar}.}
	\label{fig:allRvar}
\end{figure*}

In Figure \ref{fig:flare-sindex}, we classify solar-type stars into three subsets according to (1) whether the star generates superflares or not and (2) how much energy the superflare have released. The value $4.68 \times 10^{34}$ erg is the mean value ($\mu$) of the normal distribution fitting in Figure \ref{fig:Mflare}(a). The correlation between $S$-index and $R_{\rm var}$ is suggested to be true by \citet{2015PASJ...67...33N} and \citet{2016NatCo...711058K}. They concluded that the larger $R_{\rm var}$ may indicate that larger magnetic features (e.g. star spots) are covering on the stellar surface; meanwhile, their magnetic activities (indicated by a larger $S$-index) are active. \citet{2020ApJ...894L..11Z} used more than 2000 spectra and concluded that the asymmetry of the magnetic features plays an important role in creating photometric variability. 
In Figure \ref{fig:flare-sindex}(a), the stars without any superflares (blue circles) vaguely show this correlation, but it is not as obvious as that shown in \citet{2015PASJ...67...33N} and \citet{2016NatCo...711058K}. 
Here the unbalanced distributions of the $S$-index and $R_{\rm var}$ might be the main reason for the seeming dispersion of this correlation. From Figure \ref{fig:flare-sindex}(b) and (c), scatters are more concentrated in the area with an $S$-index ranging from 0.2 to 0.3 and $R_{\rm var}$ ranging from $10^{-3}$ to $10^{-2}$. Besides, we have 7375 nonflare stars in our database, which is more than the 1400 scatters in Figure 6 of \citet{2016NatCo...711058K}. For flare stars (red and yellow circles), the correlation is much more obvious. As we can still see the positive tendency of this correlation from nonflare stars, 
we think the magnetic features (e.g. star spots) might be the main reason for the overall positive trend of this correlation. And in small $S$-index intervals, dispersion of the $R_{\rm var}$ value may be caused by asymmetry of the magnetic features on the stellar surface. Moreover, it is also reasonable to consider inclination angle between the rotation axis and observing direction \citep[e.g.][]{2013ApJ...771..127N,2020ApJ...902...73I} as another kind of asymmetry to affect $R_{\rm var}$. Besides, we also see from Figure \ref{fig:flare-sindex}(a) that those flare stars are not with $R_{\rm var} \gtrsim10^{-1}$, while their $S$-indexes become larger. Especially for those flare stars with $S \gtrsim 0.4$, unlike what is expected according to the positive correlation between $R_{\rm var}$ and $S$-index, we do not see larger $R_{\rm var} \gtrsim 0.1$. We speculate that there seems to be a saturation-like feature at $R_{\rm var} \sim 0.1$. However, as there are only 79 flare stars that do not make this saturation-like feature statistically convincing, we will specifically discuss this saturation-like feature from Figure \ref{fig:allRvar} in the following.  

In Figures \ref{fig:flare-sindex} (b) and (c), as only 79 superflare stars have been observed by LAMOST, fitting the normal distribution to the two subsets may not be rational. We use the mean and median values of their distributions to represent the overall level of distribution, which is shown as short straight solid or dashed lines. Specially, from the $S$-index distribution, flare stars basically have higher $S$-indexes than nonflare stars. Red histograms have higher mean and median values of the $S$-index than yellow histograms, which indicates that flare stars with higher flare energies ($E_{\rm flare}>4.68\times10^{34}$ erg) tend to possess greater chromospheric activity than those flare stars with lower flare energies ($E_{\rm flare}<4.68\times 10^{34}$ erg). Apart from that, the majority of flare stars have more active chromospheric activity than the Sun.

We also see that flare stars have greater photometric variability ($R_{\rm var}$). In order to draw conclusions from more convincing results, we use all solar-type stars in this work to get a distribution of $R_{\rm var}$ as shown in Figure \ref{fig:allRvar}(a). The same classification for three subsets in Figure \ref{fig:flare-sindex} is also practiced. Overall, flare stars generally have larger $R_{\rm var}$ than nonflare stars. In Figure \ref{fig:allRvar}(a), we do not use the normal distribution to fit the red histograms, because they do not morphologically show the shape of normal distributions. However, many more flare stars in this subset are concentrated in the higher $R_{\rm var}$ range than yellow histograms. In the panel (b) of Figure \ref{fig:allRvar}, we plot the stellar $R_{\rm var}$ versus the maximum flare energy of the star. Gray squares represent mean values of $\log E_{\rm flare,max}$ (maximum flare energy in logarithm scale) in 11 $R_{\rm var}$ bins. From the gray squares, the increasing trend is much more obvious. We also combine 400 superflare stars from our previous work \citep{2020ApJ...890...46T}, and this positive tendency is also shown by gray squares in panel (d) of Figure \ref{fig:allRvar}.

\begin{figure*}[!h]
	\centering
	\includegraphics[width=0.6\linewidth]{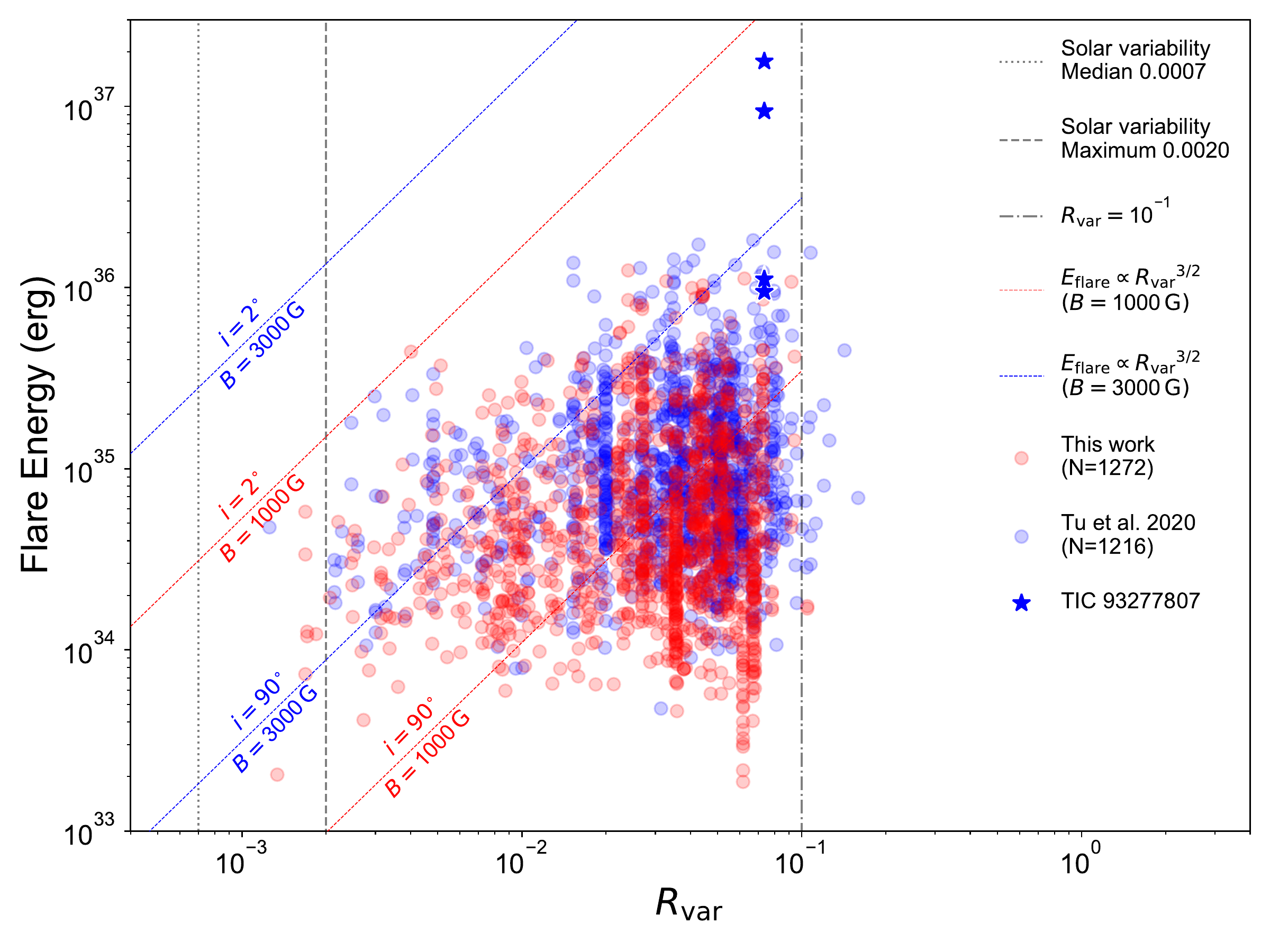}
	\caption{Diagram of all superflare energies and values of $R_{\rm var}$, which is similar to Figure \ref{fig:allRvar}(b) but includes all superflares from \cite{2020ApJ...890...46T} and this work.} 
	\label{fig:rvar-energy}
\end{figure*}

Nonflare stars generally have lower values of $S$-index and $R_{\rm var}$ than flare stars, which indicates that their magnetic activity may be less active and may not produce any energetic magnetic features (e.g. larger star spots) to generate superflares \citep{2013PASJ...65...49S}. For flare stars, their maximum superflare energy is correlated with stellar $R_{\rm var}$, i.e., the greater $R_{\rm var}$ they have, the bigger magnetic features they may host, and the more energetic a superflare can be generated. And larger magnetic features may also increase the probability of the star generating superflares, which may explain the positive correlation between flare frequency for a single star and its $R_{\rm var}$ as shown in Figure \ref{fig:compare rate}(a). Besides, the tendency of many more scatters being located on the lower right side of Figure \ref{fig:rvar-energy} can also be explained by this improved probability. Specific discussion of Figures \ref{fig:rvar-energy} and \ref{fig:compare rate}(a) can be found in Sections \ref{sec:energy and rvar} and \ref{sec:rate}, respectively. Besides, the above results  indirectly reinforce that solar-type stars can generate superflares through magnetic activity (Sections \ref{sec:frequencies} and \ref{sec:e-d-relation}), as long as their magnetic features are active enough. From the view of the possibility for the Sun to generate superflares, the above results enhance the conclusion that those flare stars are much more active than the Sun, as the flare stars in our combined database (711 superflare stars) have larger values of $R_{\rm var}$ 
than those of the Sun.

From Figure \ref{fig:allRvar}(a), we may also see that red histograms saturate around $R_{\rm var} \sim 10^{-1}$. We also draw the line in panel (b) and find that for flare stars, there is an upper limit at $R_{\rm var} \sim 10^{-1}$.
Larger flares may be generated from stars with greater $R_{\rm var}$, according to the positive relation between $R_{\rm var}$ and maximum flare energy shown in Figures \ref{fig:allRvar}(b) and (d). But those superflares are not generated from stars with $R_{\rm var} \gtrsim10^{-1}$. The 22,228 nonflare stars (blue histograms) shown in panel (a) also have extremely fewer counts over $R_{\rm var} \sim 10^{-1}$. So, this indicates that for general solar-type stars, the surface magnetic features (e.g. star spots) and their asymmetry distributions may not be able to cause $R_{\rm var}$ greater than $\sim 10^{-1}$. Furthermore, there are still a few nonflare stars with even larger $R_{\rm var}$ than $R_{\rm var} \sim 10^{-1}$, as shown in Figure \ref{fig:flare-sindex}. Potentially, they also generate superflares that are not recorded by TESS. But this possibility may not affect our statistical conclusions. It is also possible that those stars are ellipsoidal binaries, but we think that they will not take a large portion of our data set. From Kepler observations, only about 1.3\% of all Kepler targets are ellipsoidal binary systems \citep{2016AJ....151...68K}. By the way, from the $R_{\rm var}$ distributions in Figure \ref{fig:allRvar}(c), the $R_{\rm var}$ of the flare stars in this work (red histograms) is lower than that of \citet{2020ApJ...890...46T}. As we have discussed in Section \ref{sec:energy distributions}, these results may be the reason why the superflare energies of this work are slightly lower than those of \citet{2020ApJ...890...46T}, as shown in Figure \ref{fig:Mflare}(d).

Under the assumptions that we observe the star with an observing inclination angle of $90^{\circ}$, and the distribution of its magnetic features (e.g. star spots) is extremely asymmetrical, according to our results, it is hard for a solar-type star to be covered by brighter (e.g. faculae) or darker (e.g. star spots) magnetic features with an area over 10\% of the stellar surface ($R_{\rm var} \sim 10^{-1}$ and $10^{-1} \times 100$). It must be noted that, if we consider the limb-darkening law and the temperature of stellar spots as factors for measuring photometric variation \citep{2020PASJ..tmp..253M}, the actual coverage of magnetic features for some stars may be larger than 10\%. Considering the limb-darkening coefficients of \citet{2000A&A...363.1081C}, the effective stellar disk is underestimated by around 15\% - 17\%  for the solar-type stars of TESS. Temperature of stellar spots is estimated by correlation with stellar surface temperature \citep{2005LRSP....2....8B,2020PASJ..tmp..253M}. Then, the saturation-like feature is still apparent but with a little bit higher value around 15\%. From this aspect, magnetic features like star spots cannot be subtly generated or congregated even larger. So, it may prohibit the maximum energy of the superflare from getting even bigger. It may be the reason for superflare energies saturating at around $\sim 10^{37}$ erg \citep{2015ApJ...798...92W}. We also get a similar result with a limitation of flare energy around $\sim 10^{36}$ erg from the TESS observation. Superflares from \citet{2015ApJ...798...92W} are systematically with higher energies than ours. Anyway, the saturation of superflare energy may be caused by the coverage limitation of stellar magnetic active regions.

This saturation-like feature of brightness variation amplitude may not be caused by the selection effect of different telescopes. Using Kepler data, \citet{2012Natur.485..478M} and \citet{2013ApJ...771..127N} also found that the upper limit of the brightness variation amplitude of superflare stars is around $10^{-1}$. However, they mainly focused on the scale correlation between brightness variation and flare peak amplitudes and did not discuss this saturation-like feature. Furthermore, \citet{2016AJ....152..113R} used K2 data, and also showed this feature for stars from the Pleiades cluster, as shown in Figure 12 of their work.
It is necessary to figure out what physical mechanism caused this saturation-like feature. This is also an interesting topic for the study of stellar activities. A bunch of simulation works have focused on reconstructing light curves for observed stars by using artificial functions of star spots and faculae \citep[e.g.][]{2020ApJ...901...14B,2020ApJ...902...73I,2020ApJ...901L..12I,2020A&A...633A..32S}. These simulations are all based on morphologically artificial functions to describe the generation and evolution of star spots and faculae. These functions are generally without any internal physical mechanism. \citet{2016MNRAS.463.2494F} tried to figure out spot coverage of stars from Pleiades cluster using LAMOST data. Because there are not enough data with successful spot configurations, it is hard to see a clear spot coverage limitation from their results. In the future, we hope numerous observations could statistically configure out the upper limit of star-spot coverage distribution, and numerical simulations could also consider real physical patterns \citep[e.g.][]{2006ApJ...641L..73S,2014ApJ...785...90R} of star spots to give rational explanations of this saturation-like feature. 
 
 Besides, the star TIC 93277807 with $1.77 \times 10^{37}$ erg superflares has $R_{\rm var}=0.0735$, which is also less than $10^{-1}$. We show TIC 93277807 as a blue star in Figures \ref{fig:allRvar} (b) and (d). Star TIC 93277807 is far away from the majority of superflare stars. It increases the necessity for analyzing TIC 93277807 with details in the future. 
 Star TIC 93277807 deviates from other flare stars, and its energy is almost more than 1 order of magnitude higher than the others. This superflare can be treated as an extreme event. According to the Dragon-King hypothesis \citep{2009arXiv0907.4290S,2012EPJST.205....1S}, this kind of event may belong to another population, rather than the general distribution, where relatively small events (e.g. the majority of superflares) belong. The extreme event may be generated by different mechanisms, e.g. processes of synchronization and amplification. \citet{2019ApJ...880..105A} found that Dragon-King events are rare in solar and stellar flares. From their results, a single model is enough to describe small, large, and extreme flares. However, they also believed that, for solar and stellar flares, processes of synchronization and amplification may not be possible. In the future, it will also be attractive to solve what caused the most enormous superflare on TIC 93277807.
 
\begin{figure*}[!h]
	\centering
	\includegraphics[width=0.6\linewidth]{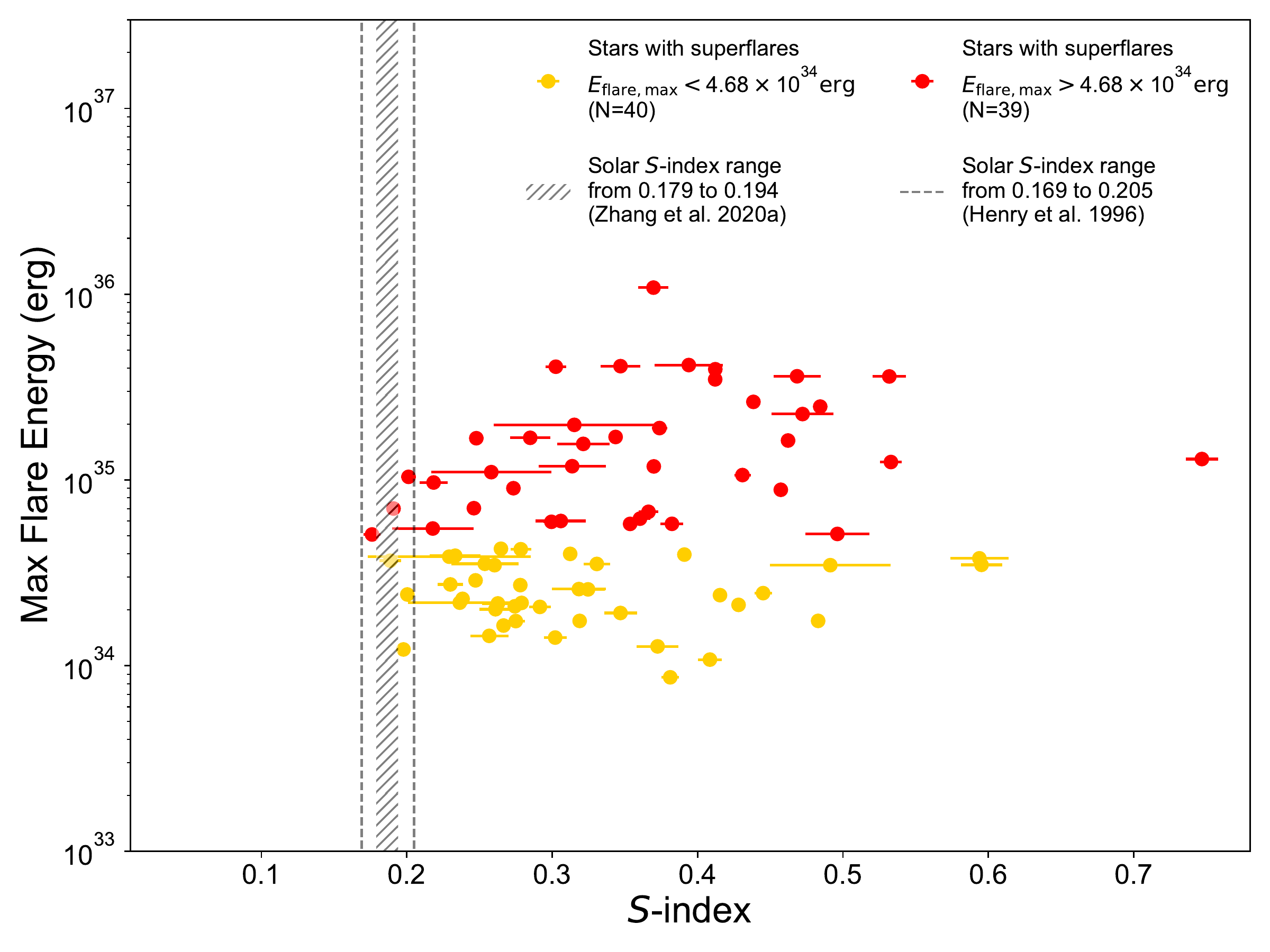}
	\caption{Scatter plot of 79 flare stars, which are measured by LAMOST. Their $S$-indexes and the maximum superflare energy of the corresponding star are included in this diagram. As in Figure \ref{fig:flare-sindex}, red and yellow represent stars separated by the energy division ($4.68\times 10^{34}$ erg). Also, the gray hatched area and two dashed vertical lines represent solar $S$-index ranges, which are imported from \citet{2020ApJ...894L..11Z} and \cite{1996AJ....111..439H}, respectively.} 
	\label{fig:sindex-maxenergy}
\end{figure*}

Apart from the above discussions from Figure \ref{fig:allRvar} of maximum superflare energy versus $R_{\rm var}$, we also make a diagram of energy and $R_{\rm var}$ for all superflares. In Figure \ref{fig:rvar-energy}, 1216 superflares from the TESS first year observation \citep{2020ApJ...890...46T} are represented by blue circles, and red circles stand for the 1272 superflares in this work. Comparing with Figures \ref{fig:allRvar}(b) and (d), the saturation-like feature of $R_{\rm var}\sim 0.1$ is still prominent in Figure \ref{fig:rvar-energy}. The superflares from one solar-type star with relatively smaller energy are vertically distributed under the maximum superflare of the star. For TIC 93277807, there were four superflares searched last year and represented by blue stars in this diagram. The second-largest superflare of TIC 93277807 is still away from the majority superflares of other solar-type stars. This result makes this star and its superflares more intriguing to deeply study in the future. We also make a diagram of maximum superflare energy versus $S$-index in Figure \ref{fig:sindex-maxenergy}. As $R_{\rm var}$ may be influenced by an inclination effect, the $S$-index seems to be a better indicator of stellar activity. But, from Figure \ref{fig:sindex-maxenergy}, we do not find any apparent correlation between $S$-index and maximum superflare energy. Only the upper limits of maximum flare energy from those scatters with $S \lesssim 0.4$ may be positively correlated with the $S$-index. A larger $S$-index indicates a more active stellar chromosphere, which may increase the possibility for the star to generate superflares. More apparently, this opinion can also be supported by the positive correlation between flare frequency for a single star and $S$-index, as shown in Figure \ref{fig:compare rate}(e). However, there are only 79 superflare stars observed by LAMOST that may not give statistically convincing results. In the future, we expect that more spectral information on superflare stars can be observed by sky survey projects to test these results.

\subsection{Superflare Energy and $R_{\rm var}$}
\label{sec:energy and rvar}
Some works have studied the correlation between star spots and energies of superflares \citep[e.g.][]{2013ApJS..209....5S,2017PASJ...69...41M,2019ApJ...876...58N,2021ApJ...906...72O}. Through assuming the relation between the size of the star spot and superflare energy, their works indicated that the same physical processes are found from stellar superflares and solar flares. Here we try to deduce a simple correlation between superflare energy and $R_{\rm var}$. The variation of normalized stellar light curves and size of star spots can be connected by
\begin{equation}\label{equ:Rvar-Aspot}
	A_{\mathrm{spot}}=\frac{\Delta F}{F} A_{\mathrm{star}}\left[1-\left(\frac{T_{\mathrm{spot}}}{T_{\mathrm{star}}}\right)^{4}\right]^{-1}
	=R_{\rm var}A_{\mathrm{star}}\left[1-\left(\frac{T_{\mathrm{spot}}}{T_{\mathrm{star}}}\right)^{4}\right]^{-1},
\end{equation}
where $\Delta {F}/{F}$ describes the amplitude variation of the star \citep[][etc.]{2019ApJ...876...58N}, which can also be replaced by the photometric variability ($R_{\rm var}$). Here $A_{\rm star}$ is the apparent stellar size ($\pi {R_{*}}^{2}$), and $T_{\rm spot}$ and $T_{\rm star}$ are the temperatures of the stellar surface and star spot, respectively. The energy of the superflare can be derived by 
\begin{equation}\label{equ:E-Aspot}
E_{\rm {flare }} \approx f E_{\rm {mag }} \approx \frac{f B^{2} L^{3}}{8 \pi} \approx \frac{f B^{2}}{8 \pi} A_{\rm {spot }}^{3 / 2},
\end{equation}
where $L$ and $B$ are the size and magnetic field strength of the star spots, respectively \citep[][etc.]{2019ApJ...876...58N}, and $f$ is the fraction of magnetic energy ($E_{\rm mag}$) that is released through superflares. 

From research of solar eruptive events, \citet{2012ApJ...759...71E} found that the bolometric energy of solar flares takes, on average, $\sim 6\%$ of the magnetic energy in the active region. Typically, $f$ is usually assumed to be on the order of $10\%$ \citep[e.g.][]{2013PASJ...65...49S,2015EP&S...67...59M,2019ApJ...876...58N}. In this work, we also take $f=0.1$ in Equation (\ref{equ:E-Aspot}). The magnetic field of spots in Equation (\ref{equ:E-Aspot}) is absolutely different from the coronal magnetic field strength in Section \ref{sec:e-d-relation} and Figure \ref{fig:dura-energy}(b). Here we take $B = 1000\,{\rm G}$ according to the idea that the typical magnetic field strength of sunspots is on the order of $1000\,{\rm G}$ \citep[e.g.][]{2013PASJ...65...49S,2015EP&S...67...59M,2019ApJ...876...58N}. We also add $B = 3000\,{\rm G}$ as an upper-limit magnetic field strength of spots \citep{2019ApJ...876...58N}.
As our database only includes solar-type stars, we assume $A_{\rm star} = \pi {R_{\odot}}^{2}$, $T_{\rm star} = 5800$ K (solar surface temperature), and $T_{\rm spot}= 4000$ K in Equation (\ref{equ:Rvar-Aspot}). By considering the inclination effect \citep[with inclination angle $i$;][]{2013ApJ...771..127N}, the correlation of $E_{\rm flare}$ and $R_{\rm var}$ can be written as
\begin{eqnarray}\label{equ:E-A-sini}
	E_{\rm {flare }} \approx 1.10 \times 10^{37}\;\left(\frac{R_{\rm var}}{\sin i}\right)^{3/2}\; {\rm erg}\;({\rm for }\;B=1000\,{\rm G}),
	\\ \nonumber 
	E_{\rm {flare }} \approx 9.86 \times 10^{37}\;\left(\frac{R_{\rm var}}{\sin i}\right)^{3/2}\; {\rm erg}\;({\rm for }\;B=3000\,{\rm G}).
\end{eqnarray}
These correlations ($E_{\rm flare}\propto R_{\rm var}^{3/2}$) are shown by red and blue dashed lines (with inclination angles labeled by $i=2^{\circ}$ and $90^{\circ}$) in Figures \ref{fig:allRvar}(b) and \ref{fig:allRvar}(d) and \ref{fig:rvar-energy}.

From Figures \ref{fig:allRvar}(d) and \ref{fig:rvar-energy}, there are many scatters obviously located on the lower right side of the plot, which is also suggested by \citet{2019ApJ...876...58N}. This tendency can be explained by the idea that a greater $R_{\rm var}$ indicates bigger magnetic features on the star, which may increase the probability for the star to generate superflares as discussed in Section \ref{sec:sindex and rvar}. As shown in Figure \ref{fig:compare rate}(a), the correlation between flare frequency for a single star and $R_{\rm var}$ may also support this opinion. Besides, from Figures \ref{fig:allRvar}(b) and (d) and \ref{fig:rvar-energy}, almost all superflares are below the line of $i=2^{\circ}$ ($B=1000\,{\rm G}$). The case of $i=2^{\circ}$ is where we observe the star from a nearly stellar pole-on direction, which is the upper limit of Equation (\ref{equ:E-A-sini}). So, from the above results, we confirm that superflare energy is basically released by the energy stored around the star spots. These results strengthen our conclusions in Section \ref{sec:sindex and rvar} that the connection between magnetic features (e.g. large star spots) and superflares is subsistent, and they also strengthen the consistency of stellar superflares and solar flares. Similar discussions can also be found in Section \ref{sec:frequencies} and \ref{sec:e-d-relation}.

\subsection{Flare Frequency of Single Star versus Other Properties}
\label{sec:rate}
\begin{figure*}[!h]
	\centering
	\includegraphics[width=1\linewidth]{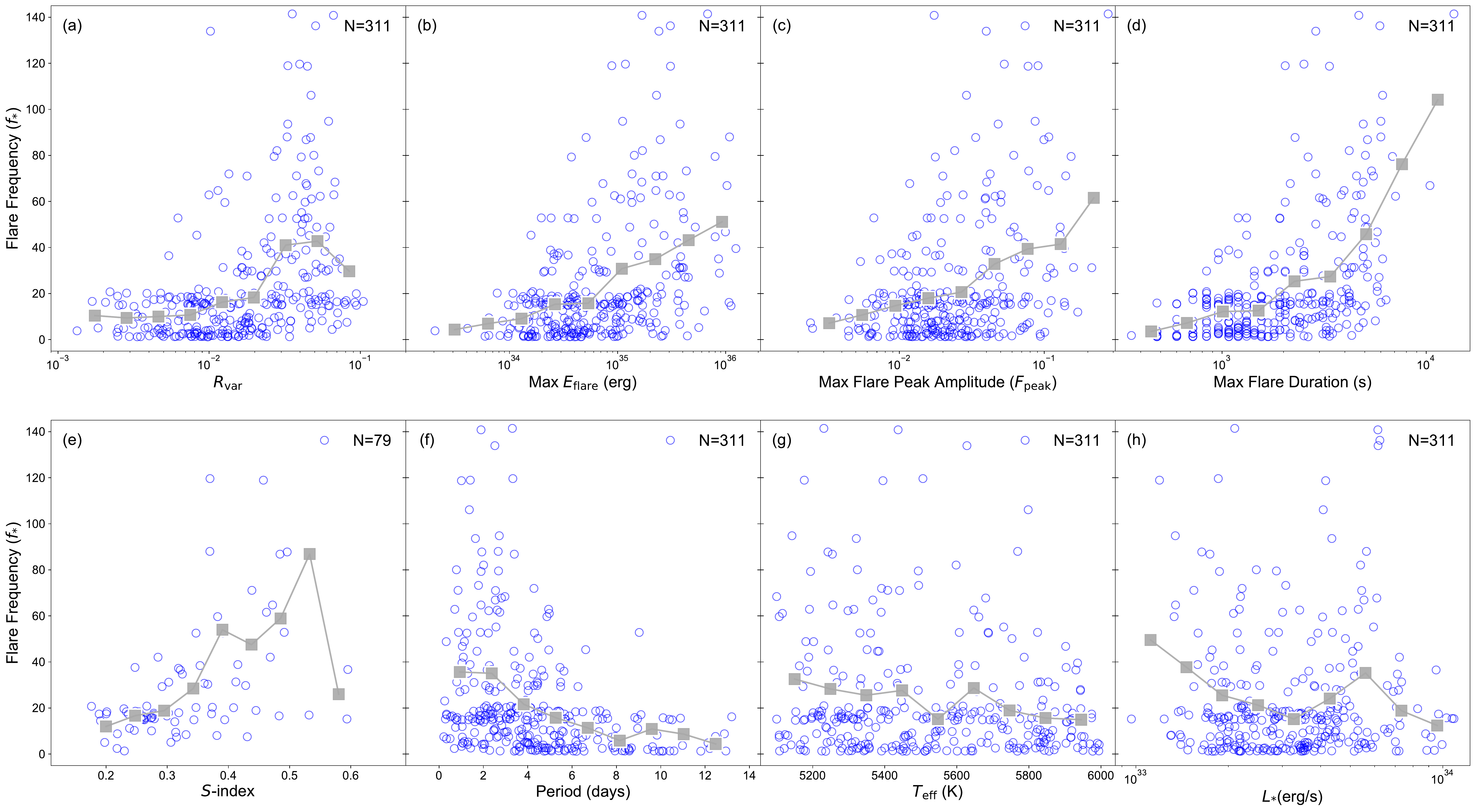}
	\caption{Plots of flare frequency ($f_{\rm *}$) versus other stellar properties. Here 311 flare stars are included, of which 79 stars have $S$-indexes. The mean value of the flare frequency as a function of different stellar properties of nine bins are marked by gray squares. Panels (a) and (e) show plots of $R_{\rm var}$ and $S$-index, respectively. Plots of the maximum flare energy($E_{\rm flare}$), peak amplitude ($F_{\rm peak}$), and duration are shown in panels (b), (c), and (d), respectively. In panels (f), (g), and (h), periods, $T_{\rm eff}$, and $L_{*}$ stand for stellar periodicity, surface temperature, and luminosity, respectively. }
	\label{fig:compare rate}
\end{figure*}

Here we study the flare frequency for a single star ($f_{*}$ from Equation (\ref{equ:activefrequency})), and try to explore the relations between $f_{*}$ and other stellar properties. Due to the unique observing mode of TESS, stars are not observed with the same time interval. As we use the effective observing time to roughly estimate $f_{*}$ through Equation (\ref{equ:activefrequency}), and many stars are just observed by one sector (Figure \ref{fig:SetN}), we just show statistically obtained relations and try to connect with empirical theorems of stellar evolution. We also use the mean flare frequency in each bin of different properties to depict their relations, as shown by gray squares in each panel of Figure \ref{fig:compare rate}.

In Figure \ref{fig:compare rate}(a), those flare stars have a  relative higher $R_{\rm var}$ in the range between $10^{-3}$ and $10^{-1}$. The mean $f_{*}$  is significantly increased with $R_{\rm var}$ higher than $10^{-2}$. Apart from that, in Figure \ref{fig:compare rate}(e), for those superflare stars with $S$-indexes, the positive relation seems more obvious, as shown by gray squares. The distribution of these flare stars are also indicating that stars with higher $S$-indexes generate superflares more frequently. 
As discussed in Sections \ref{sec:sindex and rvar} and \ref{sec:energy and rvar}, these results reinforce that magnetic activities are positively correlated with the generation of superflares. 

For Figures \ref{fig:compare rate} (b)-(d), the maximum energy ($E_{\rm flare}$), duration, and flare peak amplitude ($F_{\rm peak}$) of superflares are all collected. Significant positive relations between these three parameters and $f_{*}$ can be found not only from the gray squares but also from their scatter distribution. 
Statistically, if the star can generate a much larger superflare, it will frequently produce more relative, smaller superflares, which will apparently increase the flare frequency of the star. So, the flare frequency ($f_{*}$) for single stasr can be positively correlated with these three parameters. 

From Figure \ref{fig:compare rate} (f)-(h), we try to explore whether there are relationships between $f_{*}$ and stellar evolution. As solar-type stars evolving, it will rotate slower \citep{1984ApJ...279..763N,2003ApJ...586..464B}, and become more luminous and hotter \citep{2000MNRAS.315..543H}. We use stellar periodicity, effective temperature and luminosity as indicators of star evolution. In Section \ref{sec:frequencies}, we have found that slowly rotating stars have lower flare frequencies than rapidly rotating stars. In Figure \ref{fig:compare rate}(f), a similar result is also shown by gray squares and scatter distribution gathering in the short-period range. So, it also indicates that young stars are rapidly rotating stars, more active and with higher flare frequency. From the gray squares in panels (g) and (h), we also get slightly negative correlations between $f_{*}$ and surface temperature and luminosity. Compared with less evolved stars, it seems that the stars with larger $T_{\rm eff}$ and $L_{*}$ generate superflares less frequently. From the scatter distribution, there are some hotter and more luminous stars with greater $f_{*}$ from scatters in panels (g) and (h) of Figure \ref{fig:compare rate}, which makes the negative correlation less obvious than that in panel (f). 

\section{Summary}
\label{sec:Summary}
In this work, we select 22,539 solar-type stars, after cross-matching with catalogs from HIP2 and Gaia-EDR3. By using the improved method for visual inspection, 1272 superflares are found from 311 solar-type stars. For the first time, we combine objects of TESS together with corresponding spectra from { LAMOST}. As the TESS second-year observation is overlapped by LAMOST in the northern hemisphere of the sky, 7454 solar-type stars have spectra from LAMOST DR7 v1.1 and DR8 v0. 
For analyzing chromospheric activity and photometric variability, the $S$-index and $R_{\rm var}$ are used.

The overall energy distribution of superflares is still slightly higher than that of Kepler, which may be due to the TESS filter being more sensitive toward relatively large superflares. But the distribution is lower than the TESS first-year observation \citep{2020ApJ...890...46T}, which may be due to the improved accuracy of visual inspection and more flares have basically smaller flare peak amplitudes than those of our previous work. Generally, young solar-type stars are fast-rotating stars ($P<10$ days), which are much more active than slowly rotating stars ($P>10$ days; refer to Figure \ref{fig:oc_fre}(b)). And those stars with greater effective temperatures ($5600 \mathrm{K} \leqslant T_{\mathrm{eff}}<6000 \mathrm{K}$) generate superflares less frequently than cooler stars (with $5100
\mathrm{K} \leqslant T_{\mathrm{eff}}<5600 \mathrm{K}$; refer to Figure \ref{fig:oc_fre}(c)). From the power-law correlation $d N / d E \propto E^{\gamma}$, we get $\gamma \sim -1.76 \pm 0.11$ for all superflares. Combining superflares from TESS two-year observations and solar flares from previous literature, we find that the superflare frequency distribution can be basically treated as an extension of solar flares. Their distributions are around $\gamma \sim -1.80$, which indicates that superflares may share the same mechanism with solar flares. A higher flare frequency for the Sun-like stars of TESS is found, compared with that of Kepler (refer to Figure \ref{fig:compare flare}(b)). 

Besides, the correlation between superflare energy and duration also gives $T_{\text {duration }} \propto E^{\beta} $ with $\beta \sim 0.42 \pm 0.11$, which may still indicate that even stellar superflares share the same mechanism with solar flares. We also make plots of stellar periods versus flare energies. The upper limit of the flare energy descends with period at the tail part. Because TESS may limit the accuracy of periodic estimation due to its unique observation mode, we expect another accurate periodic estimation can be done for all TESS objects.

For the most important part of this work, we evaluate the photometric variability through $R_{\rm var}$, and chromospheric activity by $S$-index. After excluding the potential risk of using LAMOST spectra in about an 8 yr observation interval, which may not be simultaneously obtained with the TESS second-year observation, we get the $S$-index and $R_{\rm var}$ distributions in Figure \ref{fig:flare-sindex}. Apparently, the relation between $S$-index and $R_{\rm var}$ is still obvious for superflare stars. So, magnetic features (e.g. star spots) covering the stellar surface might be the main factor for causing photometric variability \citep{2015PASJ...67...33N,2016NatCo...711058K}. Apart from that, the asymmetry of magnetic feature distribution \citep[e.g.][]{2020ApJ...894L..11Z} and observing inclination angles \citep[e.g.][]{2013ApJ...771..127N,2020ApJ...902...73I} are other factors affecting the brightness variation amplitude ($R_{\rm var}$). Overall, flare stars have a higher $S$-index and $R_{\rm var}$ than the Sun, which indicates that the Sun is generally less active than a superflare star. 

Furthermore, we find that $R_{\rm var} \sim 0.1$ may be a saturation-like phenomenon of  brightness variation amplitude, which indicates that  the magnetic features covering the stellar surface may be generated or congregated even larger. Furthermore, it may prevent the superflare energy from getting larger than $\sim 10^{36}$ erg. Besides, the most energetic superflare is still the one we found in our previous work. The superflare from TIC 93277807 releases $1.77 \times 10^{37}$ erg of energy. And this target is located far from the majority of superflare stars in the $R_{\rm var}$-maximal energy plot. However, it has $R_{\rm var} \sim 0.0735$, which is less than $R_{\rm var} \sim 10^{-1}$. We speculate that this superflare can be treated as an extreme event, which is expected by the Dragon-King hypothesis. According to this hypothesis, the mechanism of this superflare may be significantly different from that of other superflares. 

The relation of $E_{\rm flare} \propto {R_{\rm var}}^{3/2}$ is applied in this work. Almost all superflares with maximum energy are lower than the line of $E_{\rm flare} \propto {R_{\rm var}}^{3/2}$ with inclination angles $i=2^{\circ}$ and a magnetic field of star spots $B=1000\,{\rm G}$. These results reinforce that the energy of the stellar superflare is the magnetic energy stored around the star spots.
 
The positive relation between flare frequency ($f_{*}$) and $S$-index $R_{\rm var}$ supports the idea that magnetic activity is positively correlated with the flare frequency of a star. This can also explain the tendency of scatters located on the lower right side of Figures \ref{fig:allRvar}(d) and \ref{fig:rvar-energy}. We also find that the larger a superflare is, the more relative smaller superflares can be generated on the same star. Besides, $f_{*}$ is negatively correlated with periodicity, surface temperature, and luminosity, which indicates that young solar-type stars are basically activ, and produce superflares more frequently.

There is no doubt that, as TESS extended observations are now ready to begin, many more superflares can be collected for statistical analyzing in the future. Besides, we also look forward to the upcoming missions, e.g., the CHaracterising ExOPlanets Satellite (CHEOPS) and PLAnetary Transits and Oscillations of stars (PLATO), that will be searching for stellar superflares. We could forecast that, with unique surveying mode of TESS observations, combining spectral information from ground-based sky surveys will definitely be beneficial for statistical works of stellar activity. 
In near future, spectral data from the Sloan Digital Sky Surveys \RNum{5}\citep[][]{2017arXiv171103234K} and 4 meter multi-object spectroscopic telescope \citep[4MOST;][]{2012SPIE.8446E..0TD} will cover the southern hemisphere of the sky. With all kinds of data gathering for statistical studies of superflare stars, we will expect to accurately describe the frequency of superflares. In the future, it will be exciting to study how stellar coronal mass ejection (CME) or prominence eruptions \citep{2019NatAs...3..742A,2019A&A...623A..49V,2020PASJ..tmp..253M} relate to superflares.

\section*{Acknowledgements}
We thank the anonymous referee for helpful and detailed suggestions and comments. We would like to thank Ji-Wei Xie, Xin Cheng, Niu Liu, He-Chao Chen, and Xiang-Song Fang for valuable discussions. This work is supported by the National Natural Science Foundation of China (grants U1831207, 12003027, and 11803012) and National Key Basic R\&D Program of China (grant 2019YFA0405500). H. F. W. is supported by the LAMOST Fellow Project, China Postdoctoral Science Foundation (grants 2019M653504 and 2020T130563), Yunnan Province Postdoctoral Directed Culture Foundation, and the Cultivation Project for LAMOST Scientific Payoff and Research Achievement of CAMS-CAS. We acknowledge the team of the Transiting Exoplanet Survey Satellite (TESS) for support in applying enormous public open resource. The Guoshoujing Telescope (the Large Sky Area Multi-Object Fiber Spectroscopic Telescope, LAMOST) is a National Major Scientific Project built by the Chinese Academy of Sciences. Funding for the project has been provided by the National Development and Reform Commission. LAMOST is operated and managed by the National Astronomical Observatories, Chinese Academy of Sciences.

\bibliography{export-bibtex}{}
\bibliographystyle{aasjournal}

\FloatBarrier
\begin{table*}[!h]
	\renewcommand{\arraystretch}{1}
	\addtolength{\tabcolsep}{+1pt}
	\caption{Properties of non-flare stars.}
	\label{tab:non-flare stars}
	\centering
	\begin{tabular}{lrccccrccc}
		\hline
		\hline
		\multicolumn{1}{c}{TESS ID} & ${T}_{\rm mag}$ $^{\rm a}$& $T_{\rm eff}$ $^{\rm b}$ & $\log{g}$ $^{\rm c}$ & Radius $^{\rm d}$   & Period $^{\rm e}$ & Set-$n$ $^{\rm f}$ & $R_{\rm var}$ $^{\rm g}$ & $S$-index $^{\rm h}$ & Flag $^{\rm i}$\\
		\multicolumn{1}{c}{}           &                     & (K)              &                & ($R_{\odot}$) & (days)      &              &                &            &          \\ \hline
		346444591                &   10.09                                      & 5774               & 4.47           & 0.98          & 6.07        & 4            & 0.003859       & -       & -              \\
		230025866                &   11.10                                    & 5907               & 4.55           & 0.91          & 5.10        & 12           & 0.006319       & -        & -            \\
		198407679                &   11.44                                     & 5822               & 4.23           & 1.30          & 6.90         & 12           & 0.007828       & 0.1676 $\pm$ 0.0033  & -\\
		229736951                &    11.06                                   & 5808               & 4.51           & 0.94          & 7.82        & 13           & 0.005902       & -          & -          \\
		199549781                &   10.67                                      & 5626               & 4.30            & 1.17          & 4.68        & 12           & 0.008157       & -         & -          \\ \hline
	\end{tabular}
	\begin{flushleft}
		\textsc{Note.}\\ {
			$^{\rm a}$ TESS magnitude from TIC v8 \citep{2019AJ....158..138S}. \\
			$^{\rm b}$ Stellar surface effective temperature in units of K. \\
			$^{\rm c}$ Stellar surface gravity in log scale.\\
			$^{\rm d}$ Stellar radius in units of $R_{\odot}$.\\
			$^{\rm e}$ Stellar periodicity in units of days.\\
			$^{\rm f}$ The $n$ of Set-$n$ represents how many sectors of the star have been observed by TESS.\\
			$^{\rm g}$ Amplitude of stellar brightness variation ($R_{\rm var}$).\\
			$^{\rm h}$ The $S$-index is the indicator of stellar chromospheric activity. \\
			$^{\rm i}$ The star may be flagged as GM21, GM42, or GB, where GM21 and GM42 stand for those stars that may contain M dwarfs less than 21$''$ or within a 21$''$ - 42 $''$ distance interval, respectively, and GB stands for those stars that contain brighter stars in distances from 21$''$ to 42$''$. \\
			(This table is available in its entirety in machine-readable form.)}\\
	\end{flushleft}
	
\end{table*}

\begin{table*}
	\renewcommand{\arraystretch}{1}
	\addtolength{\tabcolsep}{+1pt}
	\caption{Properties of flare stars.}
	\label{tab:flarestars}
	\centering
	\begin{tabular}{lrccccrrrccc}
		\hline
		\hline
		\multicolumn{1}{c}{TESS ID} & ${T}_{\rm mag}$ $^{\rm a}$ & $T_{\rm eff}$ $^{\rm b}$ & $\log{g}$ $^{\rm c}$ & Radius $^{\rm d}$   & Period $^{\rm e}$ & Set-$n$ $^{\rm f}$ & Flares $^{\rm g}$ & $f_{\rm *}$ $^{\rm h}$    & $R_{\rm var}$ $^{\rm i}$ & $S$-index $^{\rm j}$  & Flag $^{\rm k}$ \\
		& & (K)              &                & ($R_{\odot}$) & (days)      &              &             & (${\rm yr}^{-1}$) &                    &          &          \\ \hline
		\multicolumn{12}{l}{Top 10 Flare Stars Sorted by Flare Frequency $f_{\rm *}$.}                                                                                                               \\  \hline
		431107890         &      7.70                  & 5203               & 4.52           & 0.85          & 2.55        & 1            & 9           & 168.41              & 0.015910           & -        & -           \\
		620711218         &     8.73                    & 5485               & 4.51           & 0.90          & 1.50        & 1            & 8           & 161.73              & 0.047516           & 0.4121  $\pm$  0.0041& -   \\
		292402144         &     10.05                    & 5165               & 4.65           & 0.73          & 0.74        & 1            & 9           & 156.78              & 0.038577           & 0.5330  $\pm$  0.0075 & -  \\
		394030788         &       9.73                  & 5231               & 4.48           & 0.90          & 3.31        & 8            & 76          & 141.44              & 0.035478           & -        & -           \\
		46674348           &      6.56               & 5437               & 4.11           & 1.43          & 1.89        & 1            & 8           & 140.82              & 0.066455           & -          & -         \\
		75622645           &      9.64               & 5628               & 4.19           & 1.33          & 2.52        & 1            & 9           & 133.93              & 0.010205           & -        & -           \\
		53795929           &     11.04             & 5505               & 4.66           & 0.77          & 3.33        & 1            & 6           & 119.63              & 0.039722           & 0.3699 $\pm$   0.0048 & -  \\
		450325956         &     11.22                 & 5176               & 4.70           & 0.69          & 1.41        & 1            & 8           & 118.95              & 0.033181           & 0.4572  $\pm$   0.0048& -   \\
		141120071         &     9.50                  & 5395               & 4.26           & 1.19          & 1.01        & 2            & 15          & 118.73              & 0.044706           & -           & -        \\
		471000657         &     6.90                 & 5798               & 4.44           & 1.02          & 1.37        & 1            & 7           & 106.10              & 0.047232           & -            & -       \\ \hline
		\multicolumn{12}{l}{Top 10 Flare Stars Sorted by Superflare Numbers.}                                                                                                                        \\ \hline
		394030788          &     9.73                & 5231               & 4.48           & 0.90          & 3.31        & 8            & 76          & 141.44              & 0.035478           & -         & -          \\
		233131533          &     10.73               & 5493               & 4.46           & 0.96          & 2.68        & 13           & 64          & 79.53               & 0.026926           & -         & -          \\
		417735733          &      9.49               & 5298               & 4.16           & 1.32          & 2.24        & 11           & 42          & 62.25               & 0.054322           & -        & -           \\
		357032575          &    9.45                 & 5100               & 4.40           & 0.97          & 2.97        & 9            & 37          & 68.39               & 0.068007           & -           & -        \\
		235930066          &      9.05               & 5191               & 4.44           & 0.93          & 4.94        & 9            & 36          & 62.88               & 0.066875           & -          & -         \\
		441807438          &       8.81              & 5181               & 4.21           & 1.22          & 1.14        & 11           & 34          & 48.84               & 0.051551           & -         & -          \\
		420137030          &     10.12                & 5935               & 4.40           & 1.09          & 1.03        & 13           & 32          & 39.54               & 0.023989           & -        & -           \\
		428988656          &     6.72                 & 5142               & 4.63           & 0.75          & 2.72        & 5            & 31          & 94.82               & 0.061714           & -       & GM21            \\
		229480487          &       10.25             & 5409               & 4.15           & 1.35          & 1.29        & 13           & 28          & 35.06               & 0.072838           & -        & -           \\
		229640059           &     10.37             & 5334               & 4.30           & 1.12          & 3.54        & 12           & 24          & 32.75               & 0.050974           & -        & -      \\ \hline    
	\end{tabular}
	\begin{flushleft}
		\textsc{Note.}\\ {Same as Table \ref{tab:non-flare stars} but for flare stars. We set this table in two parts. The first part lists the top 10 flare stars sorted by flare frequency ($f^{*}$). The second part lists the top 10 flare stars sorted by superflare number. Columns flagged as a, b, c, d, d, f, i, j and k are same with those in Table \ref{tab:non-flare stars}.\\
			$^{\rm g}$ Counts of superflares of each flare star. \\
			$^{\rm h}$ Flare frequency from Equation \ref{equ:activefrequency}. \\
			(This table is available in its entirety in machine-readable form.) 
		}\\
		\vspace{0.1em}
	\end{flushleft}
	
\end{table*}

\begin{table*}[!h]
	\renewcommand{\arraystretch}{1}
	\addtolength{\tabcolsep}{+1pt}
	\caption{Properties of superflares}
	\label{tab:Flares}
	\centering
	\begin{tabular}{ccccr}
		\hline
		\hline
		\multicolumn{1}{c}{TESS ID} & Peak Date $^{\rm a}$       & Peak Luminosity $^{\rm b}$ & Energy $^{\rm c}$ & \multicolumn{1}{c}{Duration $^{\rm d}$} \\
		\multicolumn{1}{l}{}           & \multicolumn{1}{l}{} & (erg s$^{-1}$) & (erg)       & \multicolumn{1}{c}{(s)}           \\ \hline
		420137030                      & 1840.6839            & 6.99E+32       & 1.24E+36    & 4919.96                           \\
		284358867                      & 1910.0124            & 6.26E+32       & 1.12E+36    & 5040.04                           \\
		85487971                       & 1797.3539            & 6.07E+32       & 1.08E+36    & 6000.04                           \\
		55752857                       & 1923.5225            & 4.39E+32       & 1.07E+36    & 5999.85                           \\
		364075921                      & 1782.5163            & 4.96E+32       & 1.03E+36    & 5400.02                           \\
		387020248                      & 1817.2908            & 3.09E+32       & 9.95E+35    & 5760.10                           \\
		364075921                      & 1967.7843            & 1.49E+32       & 9.20E+35    & 10439.97                          \\
		364075921                      & 1799.4219            & 2.99E+32       & 9.13E+35    & 5879.99                           \\
		364075921                      & 1811.4746            & 4.86E+32       & 8.95E+35    & 5879.96                           \\
		364075921                      & 2024.7556            & 1.84E+32       & 8.89E+35    & 9720.15                   \\ \hline       
	\end{tabular}
	\begin{flushleft}
		\textsc{Note.}\\ {
			$^{\rm a}$ Superflare peak times of TESS observation time stamps. \\
			$^{\rm b}$ The flare peak luminosity is in units of erg s$^{-1}$, which can be deduced by Equation (\ref{equ:normal distribution}), where $L_{*}$ is the stellar luminosity calculated by Equation \ref{equ:lumi}, and $F_{\rm flare}$ is the flare peak amplitude deduced by Equation \ref{equ:flare_flux} at the peak time.\\
			$^{\rm c}$ Superflare total energy.\\
			$^{\rm d}$ Superflare duration.\\
			(This table is available in its entirety in machine-readable form.)}\\
	\end{flushleft}
	
\end{table*}

\begin{table*}[!h]
	\renewcommand{\arraystretch}{1}
	\addtolength{\tabcolsep}{+10pt}
	\caption{Count distribution as a function of stellar periodicity.}
	\label{tab:num_period}
	\centering
	\begin{tabular}{cccc}
		\hline \hline
		$\log {P}$ & Solar-type Stars & Superflares & Flare Stars \\ \hline
		-1.00 & 511 & 0 & 0 \\
		-0.81 & 477 & 1 & 1 \\
		-0.62 & 338 & 14 & 6 \\
		-0.43 & 333 & 8 & 2 \\
		-0.24 & 390 & 53 & 12 \\
		-0.05 & 650 & 163 & 21 \\
		0.14 & 792 & 134 & 36 \\
		0.33 & 1763 & 483 & 59 \\
		0.52 & 5243 & 261 & 81 \\
		0.71 & 6468 & 113 & 60 \\
		0.90 & 4614 & 38 & 29 \\
		1.10 & 928 & 4 & 4 \\
		1.29 & 12 & 0 & 0 \\
		1.48 & 3 & 0 & 0 \\
		1.67 & 2 & 0 & 0 \\
		1.86 & 13 & 0 & 0 \\
		2.05 & 2 & 0 & 0\\ \hline
		Total & 22,539 & 1272 & 331 \\ \hline
	\end{tabular}
	\begin{flushleft}
		\textsc{Note.}\\ {The distribution of solar-type stars, superflares, and flare stars in 17 period bins, which are represented in log scale as $\log \rm{P}$.}\\
		\vspace{1.5em}
	\end{flushleft}
	
\end{table*}

\begin{table*}[!h]
	\renewcommand{\arraystretch}{1}
	\addtolength{\tabcolsep}{+0pt}
	\caption{Number fractions of solar-type stars}
	\label{tab:N10result}
	\centering
	\begin{tabular}{ccclcc}
		\hline \hline
		\multirow{2}{*}{Data Set}                       & \multicolumn{2}{c}{$5100\, \mathrm{K} \leqslant T_{\mathrm{eff}}<5600\, \mathrm{K}$}                            &  & \multicolumn{2}{c}{$5600\, \mathrm{K} \leqslant T_{\mathrm{eff}}<6000\, \mathrm{K}$}                            \\ \cline{2-3} \cline{5-6}
		& $N_{\rm star}(P < 10\; { \rm days})/N_{\rm all}$ & $N_{\rm star}(P \geqslant 10\; { \rm days})/N_{\rm all}$ &  & $N_{\rm star}(P < 10\; { \rm days})/N_{\rm all}$ & $N_{\rm star}(P \geqslant 10\; { \rm days})/N_{\rm all}$ \\ \hline
		Gyrochronology relation$^{\rm a}$                     & $5.1\%-7.1\%$                                    & $92.9\%-94.9\%$                                          &  & $7.1\%-10.9\%$                                   & $89.1\%-92.9\%$                                          \\
		\citet{2019ApJ...876...58N}$^{\rm b}$ & $14.1\%$                                         & $85.9\%$                                                 &  & $21.7\%$                                         & $78.3\%$                                                 \\
		\citet{2020ApJ...890...46T}$^{\rm c}$                                      & $82.6\%$                                         & $17.4\%$                                                 &  & $87.6\%$                                         & $12.4\%$                                                 \\
		This work$^{\rm d}$                                        & $80.7\%$                                         & $19.3\%$                                                 &  & $88.0\%$                                         & $12.0\%$                                                 \\ \hline
		\multicolumn{1}{l}{}                            & \multicolumn{1}{l}{}                             & \multicolumn{1}{l}{}                                     &  & \multicolumn{1}{l}{}                             & \multicolumn{1}{l}{}
	\end{tabular}
	\begin{flushleft}
		\textsc{Notes.}\\ {
			Fractions of slowly and rapidly rotating stars, which are denoted as $N_{\rm star}/N_{\rm all}$. Solar-type stars are classified according to stellar surface temperatures and stellar periods.\\
			$^{\rm a}$ Results from empirical gyrochronology relation \citep{2020ApJ...890...46T}. \\
			$^{\rm b}$ Results of Table 9 from \citet{2019ApJ...876...58N}. \\
			$^{\rm c}$ Results of Table 8 from \citet{2020ApJ...890...46T}. \\
			$^{\rm d}$ Results of this work.
		}\\
		\vspace{0.1em}
	\end{flushleft}
	
\end{table*}

\begin{table*}[!h]
	\renewcommand{\arraystretch}{1}
	\addtolength{\tabcolsep}{+1pt}
	\caption{Planet candidates of flare stars.}
	\label{tab:planets}
	\centering
	\begin{tabular}{llcccl}
		\hline \hline
		\multicolumn{1}{c}{Host Star} & \multicolumn{1}{c}{Planet ID} & \multicolumn{1}{c}{Period $^{\rm a}$} & Radius $^{\rm b}$& Equilibrium Temp. $^{\rm c}$ & \multicolumn{1}{c}{Information $^{\rm d}$} \\
		\multicolumn{1}{c}{TESS ID} & \multicolumn{1}{c}{} & \multicolumn{1}{c}{(Days)} & ($R_\oplus$) & ($K$) & \multicolumn{1}{c}{} \\ \hline
		236714379 & TOI 1254.01 & 1.0180 $\pm$ 0.000006& 9.66 $\pm$ 3.60& 1801.33 & TESS Object of Interest \\ \hline
		350132371 & TOI 1425.01 & 1.0319 $\pm$ 0.000022 & 18.96  $\pm$ 4.56 & 2249.14 & TESS Object of Interest \\ \hline
		138968089 & TOI 1429.01 & 0.641256  $\pm$ 0.000109 & 4.84 $\pm$ 7.96 & 2573.96 & TESS Object of Interest \\ \hline
	\end{tabular}
	\begin{flushleft}
		\textsc{Note.}\\ {Results of cross-matching flare stars with ExoFOP-TESS \footnote{https://exofop.ipac.caltech.edu/tess/}. Objects TOI 1254.01 and TOI 1425.01 are two hot Jupiter candidates. ObjectTOI 1429.01 is a USP planet candidate. \\
			$^a$ Periodicity of planet orbits in units of days.\\
			$^b$ Planet radius in units of Earth radius ($R_{\oplus}$).\\
			$^c$ Theoretically estimated equilibrium temperature heated by the hosting star. The original catalog does not include errors.\\
			$^d$ Other information about this planet candidate. Here these stars are only objects of interest and not yet confirmed. }\\
		\vspace{0.1em}
	\end{flushleft}
\end{table*}

\begin{sidewaystable*}[!h]
	\caption{Counts of solar-type stars, flare stars and superflares in each Set-$n$.}
	\label{tab:SetN}
	\centering
	\begin{tabular}{crrrrrrrrrrrrrrrrrrr}
		\hline \hline
		\multirow{3}{*}{Set-$n$} & \multicolumn{7}{c}{$5100\, \mathrm{K} \leqslant T_{\mathrm{eff}}<5600\, \mathrm{K}$} & \multicolumn{1}{c}{} & \multicolumn{7}{c}{$5600\, \mathrm{K} \leqslant T_{\mathrm{eff}}<6000\, \mathrm{K}$} & \multicolumn{1}{c}{} & \multicolumn{3}{c}{\multirow{2}{*}{Total}} \\ \cline{2-8} \cline{10-16}
		& \multicolumn{3}{c}{$P<10$ days} & \multicolumn{1}{c}{} & \multicolumn{3}{c}{$P>10$ days} & \multicolumn{1}{c}{} & \multicolumn{3}{c}{$P<10$ days} & \multicolumn{1}{c}{} & \multicolumn{3}{c}{$P>10$ days} & \multicolumn{1}{c}{} & \multicolumn{3}{c}{} \\ \cline{2-4} \cline{6-8} \cline{10-12} \cline{14-16} \cline{18-20}
		& \multicolumn{1}{c}{$N_{\rm star}$} & \multicolumn{1}{c}{$N_{\rm flare}$} & \multicolumn{1}{c}{$N_{\rm fstar}$} & \multicolumn{1}{c}{} & \multicolumn{1}{c}{$N_{\rm star}$} & \multicolumn{1}{c}{$N_{\rm flare}$} & \multicolumn{1}{c}{$N_{\rm fstar}$} & \multicolumn{1}{c}{} & \multicolumn{1}{c}{$N_{\rm star}$} & \multicolumn{1}{c}{$N_{\rm flare}$} & \multicolumn{1}{c}{$N_{\rm fstar}$} & \multicolumn{1}{c}{} & \multicolumn{1}{c}{$N_{\rm star}$} & \multicolumn{1}{c}{$N_{\rm flare}$} & \multicolumn{1}{c}{$N_{\rm fstar}$} & \multicolumn{1}{c}{} & \multicolumn{1}{c}{$N_{\rm star}$} & \multicolumn{1}{c}{$N_{\rm flare}$} & \multicolumn{1}{c}{$N_{\rm fstar}$} \\ \hline
		1  & 4741 & 176 & 67  &  & 965  & 3  & 3  &  & 6398  & 99  & 51  &  & 761  & 1 & 1 &  & 12,865 & 279  & 122 \\
		2  & 1295 & 59  & 22  &  & 359  & 1  & 1  &  & 1717  & 15  & 11  &  & 288  & 1 & 1 &  & 3659  & 76   & 35  \\
		3  & 662  & 38  & 15  &  & 183  & 4  & 2  &  & 877   & 38  & 18  &  & 117  & 0 & 0 &  & 1839  & 80   & 35  \\
		4  & 393  & 39  & 9   &  & 87   & 0  & 0  &  & 470   & 33  & 8   &  & 66   & 0 & 0 &  & 1016  & 72   & 17  \\
		5  & 172  & 73  & 10  &  & 49   & 0  & 0  &  & 212   & 2   & 1   &  & 37   & 0 & 0 &  & 470   & 75   & 11  \\
		6  & 105  & 48  & 6   &  & 31   & 0  & 0  &  & 135   & 40  & 5   &  & 27   & 0 & 0 &  & 298   & 88   & 11  \\
		7  & 59   & 1   & 1   &  & 25   & 0  & 0  &  & 100   & 1   & 1   &  & 21   & 0 & 0 &  & 205   & 2    & 2   \\
		8  & 53   & 83  & 4   &  & 19   & 0  & 0  &  & 87    & 1   & 1   &  & 9    & 0 & 0 &  & 168   & 84   & 5   \\
		9  & 48   & 82  & 7   &  & 17   & 2  & 1  &  & 61    & 1   & 1   &  & 10   & 0 & 0 &  & 136   & 85   & 9   \\
		10 & 54   & 2   & 1   &  & 16   & 0  & 0  &  & 81    & 2   & 2   &  & 14   & 0 & 0 &  & 165   & 4    & 3   \\
		11 & 122  & 114 & 10  &  & 37   & 0  & 0  &  & 158   & 5   & 2   &  & 27   & 0 & 0 &  & 344   & 119  & 12  \\
		12 & 263  & 64  & 13  &  & 92   & 3  & 3  &  & 387   & 63  & 14  &  & 70   & 1 & 1 &  & 812   & 131  & 31  \\
		13 & 204  & 137 & 11  &  & 69   & 3  & 2  &  & 247   & 36  & 4   &  & 42   & 1 & 1 &  & 562   & 177  & 18  \\ \hline
		Total & 8171 & 916 & 176 &  & 1949 & 16 & 12 &  & 10,930 & 336 & 119 &  & 1489 & 4 & 4 &  & 22,539 & 1272 & 311 \\ \hline
	\end{tabular}
	\begin{flushleft}
		\textsc{Note.}\\ {Set-$n$ indicates the number of sectors in which the star has been observed. According to stellar surface temperature and periodicity, the number of solar-type stars ($N_{\rm star}$), superflares ($N_{\rm flar}$), and flare stars ($N_{\rm fstar}$) for different Set-$n$ are all listed here.
		}\\
		\vspace{1.5em}
	\end{flushleft}
\end{sidewaystable*}

\end{document}